\documentclass[draftclsnofoot,onecolumn,12pt]{IEEEtran}

\usepackage{verbatim}
\usepackage{amsfonts}
\usepackage{amssymb}
\usepackage{stfloats}
\usepackage{cite}
\usepackage{graphicx}
\usepackage{psfrag}
\usepackage{amsmath}
\usepackage{array}
\usepackage{epstopdf}
\usepackage{authblk}
\usepackage{graphicx} 
\usepackage{amsthm} 
\usepackage{lipsum}
\usepackage{verbatim} 
\usepackage{authblk}
\usepackage{mathtools}
\usepackage{cuted}
\usepackage[lined,boxed,ruled]{algorithm2e}
\usepackage{booktabs}

\ifCLASSOPTIONcompsoc
\usepackage[caption=false,font=normalsize,labelfon
t=sf,textfont=sf]{subfig}
\else
\usepackage[caption=false,font=footnotesize]{subfig}
\fi

\usepackage[usenames]{color}

\newtheorem{lemma}{Lemma}

\newtheorem{proposition}{Proposition}

\def\proof{\noindent{\emph{Proof:} }}

\def\phi{\varphi}

\def\({\left(}
\def\){\right)}

\def\b0{{\mathbf{0}}}

\makeatletter
\newcommand{\removelatexerror}{\let\@latex@error\@gobble}
\makeatother

\begin{document}


\title{\huge Task-Oriented Sensing, Computation, and Communication Integration for Multi-Device Edge AI}
\author{Dingzhu Wen, Peixi Liu, Guangxu Zhu, Yuanming Shi, Jie Xu, Yonina C. Eldar, and Shuguang Cui     \thanks{\setlength{\baselineskip}{13pt} \noindent D. Wen is with Network Intelligence Center, School of Information Science and Technology, ShanghaiTech University, Shanghai, China (e-mail: wendzh@shanghaitech.edu.cn), and was with Shenzhen Research Institute of Big Data, Shenzhen, China. (Corresponding author: G. Zhu)

P. Liu is with State Key Laboratory of Advanced Optical Communication Systems and Networks, School of Electronics, Peking University, China, and Shenzhen Research Institute of Big Data, Shenzhen, China (e-mail: liupeixi@pku.edu.cn).

G. Zhu is with Shenzhen Research Institute of Big Data, Shenzhen, China (e-mail: gxzhu@sribd.cn). 

Y. Shi is  with Network Intelligence Center, School of Information Science and Technology, ShanghaiTech University, Shanghai, China (e-mail: shiym@shanghaitech.edu.cn). 

J. Xu is with the School of Science and Engineering (SSE) and the Future Network of Intelligence Institute (FNii), The Chinese University of Hong Kong (Shenzhen), Shenzhen, China (e-mail: xujie@cuhk.edu.cn). 

Y. C. Eldar is with Weizmann Institute of Science, Rehovot, Israel (e-mail: yonina.eldar@weizmann.ac.il).

S. Cui is with the School of Science and Engineering (SSE) and the Future Network of Intelligence Institute (FNii), The Chinese University of Hong Kong (Shenzhen), and Shenzhen Research Institute of Big Data, Shenzhen, China. He is also with Peng Cheng Laboratory (e-mail: shuguangcui@cuhk.edu.cn).

} }

\maketitle

\vspace{-2em}
\begin{abstract}
    This paper studies a new multi-device edge \emph{artificial-intelligent} (AI) system, which jointly exploits the AI model split inference and \emph{integrated sensing and communication} (ISAC) to enable low-latency intelligent services at the network edge. In this system, multiple ISAC devices perform radar sensing to obtain multi-view data, and then offload the quantized version of extracted features to a centralized edge server, which conducts model inference based on the cascaded feature vectors. Under this setup and by considering classification tasks, we measure the inference accuracy by adopting an approximate but tractable metric, namely discriminant gain, which is defined as the distance of two classes in the Euclidean feature space under normalized covariance.  To maximize the discriminant gain, we first quantify the influence of the sensing, computation, and communication processes on it with a derived  closed-form expression. Then, an end-to-end task-oriented resource management approach is developed  by integrating the three processes into a joint design. This \emph{integrated sensing, computation, and communication} (ISCC) design approach, however, leads to a challenging non-convex optimization problem, due to the complicated form of discriminant gain and the device heterogeneity in terms of channel gain, quantization level, and generated feature subsets. Remarkably, the considered non-convex problem can be \emph{optimally} solved based on the \emph{sum-of-ratios} method. This gives the optimal ISCC scheme, that  jointly determines the transmit power and time allocation at multiple devices for sensing and communication, as well as their quantization bits allocation for computation distortion control. 
    By using human motions recognition as a concrete AI inference task, extensive experiments are conducted to verify the performance of our derived optimal ISCC scheme.
\end{abstract}

\section{Introduction}
Edge \emph{artificial intelligence} (AI) has emerged as a promising technique to support a variety of intelligent applications, such as Metaverse and auto-driving, at the network edge \cite{letaief2021edge,zhu2020toward,park2019wireless,shi2020communication,chen2021communication,wen2022federated}. To enable these intelligent services, it is desirable to deploy well-trained machine learning models and utilize their inference capability for making decisions. This leads to a new research paradigm called edge AI model inference, or \emph{edge inference}  \cite{chen2021distributed,wang2020convergence}. 

Several techniques have been proposed for efficient implementation of edge inference. The first is called \emph{on-device inference} (see e.g., \cite{han2015deep,howard2017mobilenets,chen2019deep,lee2021decentralized}), in which the inference task is implemented at resource-limited devices. To alleviate the computation loads, in on-device inference we need to design dedicated light models such as MobileNets, or compress the deep models to reduce their sizes by e.g., pruning and quantization. However, as there are various AI tasks with many different models, this technique still has heavy storage and computation cost. To address this challenge, the technique of \emph{on-server inference} has been suggested (see e.g., \cite{yang2020energy,hua2021reconfigurable}). In this scheme, edge devices upload the input data to an edge server, which performs the model inference and sends the results back to the devices. Although on-server inference can significantly alleviate the hardware requirements of the devices, they are prone to data privacy leakage. To tackle this issue, the technique of \emph{split inference} is proposed, which splits the AI model into two submodels (see e.g., \cite{shi2019improving, huang2020dynamic, li2019edge, shao2020communication, jankowski2020joint, shao2021learning, lan2021progressive, yan2022optimal,liu2022resource}), one deployed at the devices for feature extraction, e.g. \emph{principle component analysis} (PCA) and convolutional layers, and the other at the edge server for the remaining inference task. As a result, split inference can preserve privacy by avoiding raw data transmission and reduce the hardware requirements at edge devices by offloading heavy computation loads to the edge server. Here we focus on the split inference technique to exploit these advantages.

\begin{figure}
	\centering
	\includegraphics[width=0.8\textwidth]{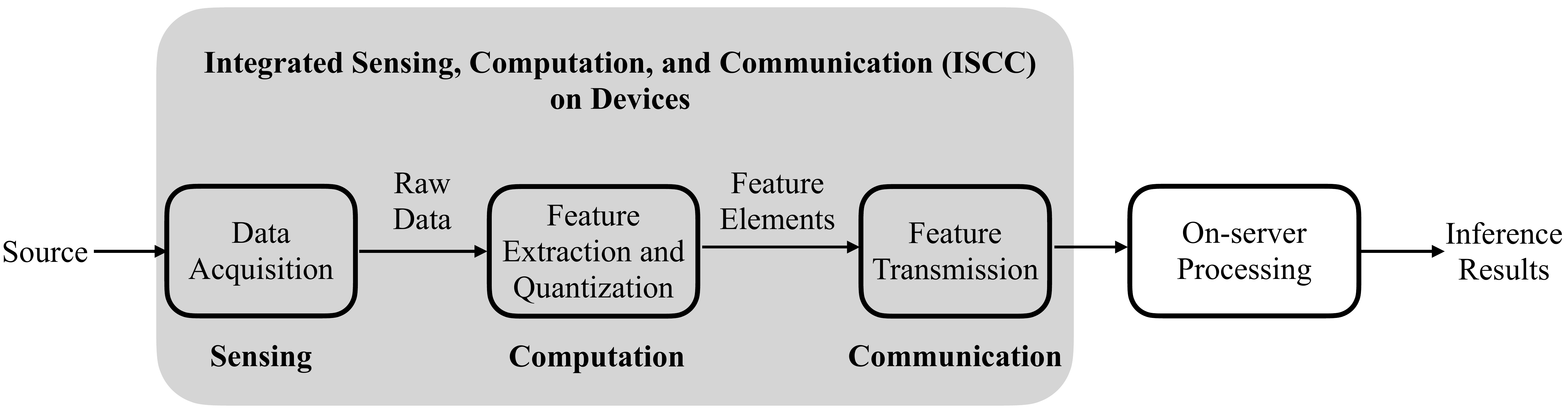}
	\caption{Integrated Sensing, Computation, and Communication (ISCC) in Edge AI Inference.}
	\vspace{-0.5cm}
	\label{fig:integration}
\end{figure}
Existing designs for split inference (see e.g., \cite{shi2019improving, huang2020dynamic, li2019edge, shao2020communication, jankowski2020joint, shao2021learning, lan2021progressive, yan2022optimal,liu2022resource}) mainly focus on reducing the devices' overhead on computation or communication. However, the workflow of split inference consists of three key processes including sensing, computation, and communication, and its full potential can hardly be unleashed by optimization from a single perspective. This thus calls for a joint design from a systematic view integrating sensing, computation, and communication. As shown in Fig. \ref{fig:integration}, the accuracy of split inference depends on the input feature vector's distortion level arising from three processes, i.e., data acquisition (sensing), feature extraction and quantization (computation), and feature transmission to edge server (communication). Particularly, sensing and communication compete for radio resources  \cite{liu2022integrated,cui2021integrating}, and the allowed communication resource further determines the required quantization (distortion) level such that the quantized features can be transmitted reliably to the edge server under a delay constraint. Thereby the three processes are highly coupled and need to be jointly considered. Furthermore, the implementation of \emph{integrated sensing, communication and computation} (ISCC) should be designed under a new \emph{task-oriented principle} that concerns the successful completion of the subsequent inference task \cite{xi2021bilimo,neuhaus2021task}. In the context of split inference, the performance metric of interest for the system is no longer throughput, but inference accuracy and latency. 
 Therefore, a real-time inference-task-oriented ISCC scheme should maximize the inference accuracy by jointly designing sensing, quantization, and transmission, under constraints on low latency and on-device resources.

To achieve task-oriented ISCC design, the employment of the recently proposed technique, called \emph{integrated sensing and communication} (ISAC), is essential as it allows efficient sensing data acquisition and feature offloading with a shared hardware \cite{ma2021spatial}. The efficiency comes from the potentially smaller form factor of the devices due to the use of shared hardware for dual functions, and better management of the shared radio resources like power and bandwidth \cite{liu2022integrated}. As one of the key potential techniques in 6G networks, ISAC has been widely studied in the existing literature, for example, optimal waveform designs for \emph{dual functional radar-and-communication} (DFRC) systems in \cite{liu2018toward,ma2020joint,pritzker2022transmit}, the beamforming designs for ISAC systems in \cite{he2022ris} and \cite{hua2021optimal}, the ISAC assisted \emph{orthogonal time frequency space} (OTFS) modulation for vehicular networks in \cite{yuan2021integrated}, and the integration of ISAC with over-the-air computation in \cite{li2022integrated}. In the aforementioned systems, sensing and communication are designed for separate goals: sensing targets obtaining high-quality localization data and communication aims at throughput maximization. However, in the context of edge AI, sensing (data acquisition) and communication (feature transmission) work together for a common goal, i.e., improving edge AI performance.

Several pioneering works investigated task-oriented ISAC schemes for edge AI. For instance, an ISAC based centralized learning system was proposed in \cite{zhang2021accelerating}, which accelerates the learning process by generating and uploading as many training data as possible from the sensing devices to the edge server. The authors in \cite{liu2022vertical} proposed a vertical federated learning based ISAC system for human motions recognition. However, the prior works above fall short in ignoring the influence of computation, and focusing only on the training phase that can usually be performed in an offline way. There still remains an uncharted area for task-oriented ISCC targeting edge inference, thus motivating the main theme of the current work.

\begin{figure}
	\centering
	\includegraphics[width=0.5\textwidth]{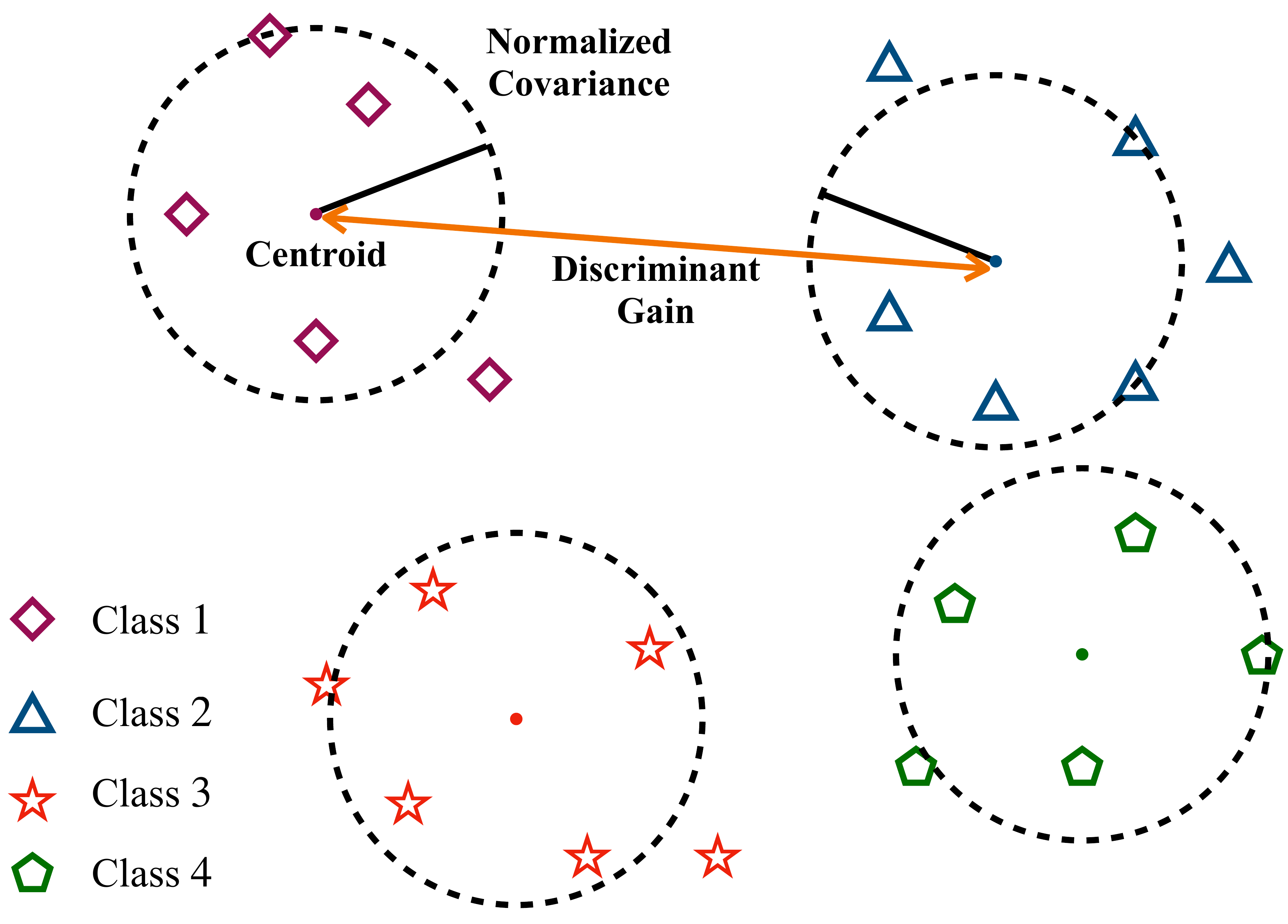}
	\caption{Geometry of discriminant gain in the feature space.}
	\vspace{-1cm}
	\label{fig:geometry}
\end{figure}
In this paper, we consider a multi-view ISAC based edge inference system with classification tasks. There are one mobile edge server (e.g., vehicle) and multiple ISAC devices equipped with DFRC systems. In this system, multiple ISAC devices perform radar sensing to obtain multi-view sensing data, and then offload the quantized version of extracted features to a centralized edge server, which conducts the model inference based on the cascaded feature vectors. The objective of this system is to maximize the inference accuracy  in a real-time manner, i.e., completing the task under a latency constraint.  Efficient implementation of the considered edge inference system relies on the design of ISCC, which faces the following technical challenges. The first main difficulty is the lack of tractable measures for inference accuracy. To address this issue, we adopt a new metric for classification tasks called  \emph{discriminant gain}, which is proposed in \cite{lan2021progressive}  and derived from the well known \emph{Kullback-Leibler} (KL) divergence \cite{kullback1951information}. The discriminant gain measures the discernibility between two classes in the Euclidean feature space, as shown in Fig. \ref{fig:geometry}. Specifically, the geometric interpretation of discriminant gain between two classes is the distance between the corresponding two classes in the feature space under normalized feature covariance. Thereby, with larger discriminant gain, the classes can be better differentiated, which leads to larger inference accuracy. As discriminant gain can provide theoretical guidance for enhancing inference accuracy, it is adopted in this work as an approximate but tractable measure. However, maximizing the discriminant gain still faces challenges arising from its complicated form of covariance normalized distance, as well as the coupling among sensing, computation, and communication, and the device heterogeneity in terms of channel gain, quantization level, and the feature elements' importance. 

To address the challenges above, a non-convex inference accuracy maximization problem is formulated under the constraints of limited on-device resources and low-latency requirement. We then propose an \emph{optimal} ISCC scheme, based on the \emph{sum-of-ratios} method, to jointly determine the transmit power and time allocation at multiple devices for sensing and communication, as well as their quantization bits allocation for computation distortion control.  To the best of our knowledge, this work represents the first attempt to design task-oriented ISCC schemes for edge AI inference systems.
The detailed contributions of this work are summarized as follows:
\begin{itemize}
\item {\bf ISCC based Edge Inference System}: A multi-view radar sensing based system is established for real-time inference tasks with concrete modeling of the sensing, computation, and communication processes. Under the system settings, we quantify the influence of sensing noise, quantization distortion, and communication capacity on the inference accuracy measured  discriminant gain with a derived closed-form expression. 

\item {\bf Inference Accuracy Maximization via ISCC Design}: Targeting maximizing the inference accuracy measured by discriminant gain, an ISCC design problem that concerns joint allocation of sensing and transmit power, communication time, and quantization bits is formulated. We then show that this problem can be  transformed into an equivalent problem with the objective being the sum of multiple quasi-linear ratios, subject to a set of convex constraints. 

\item{\bf Sum-of-ratios based Optimal Solution}:
We adopt the method of sum-of-ratios to optimally solve the reformulated problem in an iterative manner. In each iteration, a convex problem is solved, which minimizes the sum of weighted sensing and quantization distortion under given discriminant gains of class pairs. Then, the discriminant gains are updated using the previously solved distortion level of sensing and quantization. 

\item{\bf Performance Evaluation}:
Extensive simulations over a high-fidelity wireless sensing simulator proposed in \cite{Li2021SPAWC} are conducted to evaluate the performance of our proposed ISCC scheme by considering a concrete task of multi-view human motion recognition with two inference models, i.e., \emph{support vector machine} (SVM) and \emph{multi-layer perception} (MLP) neural network, respectively. It is shown that maximizing the discriminant gain is effective in maximizing the inference accuracy for both models with SVM and MLP neural networks. It is also shown that the proposed optimal ISCC scheme achieves significantly higher inference accuracy than the benchmark schemes, where sensing, quantization, and communication are separately designed or partially optimized. The superiority of multi-view inference over single-view inference is also validated.
\end{itemize}


\section{System Model}
In this section, the models of network, radar sensing and feature generation, quantization, and the metric of AI model inference accuracy are introduced.

\subsection{Network Model}
\begin{figure}
\centering
\includegraphics[width=0.8\textwidth]{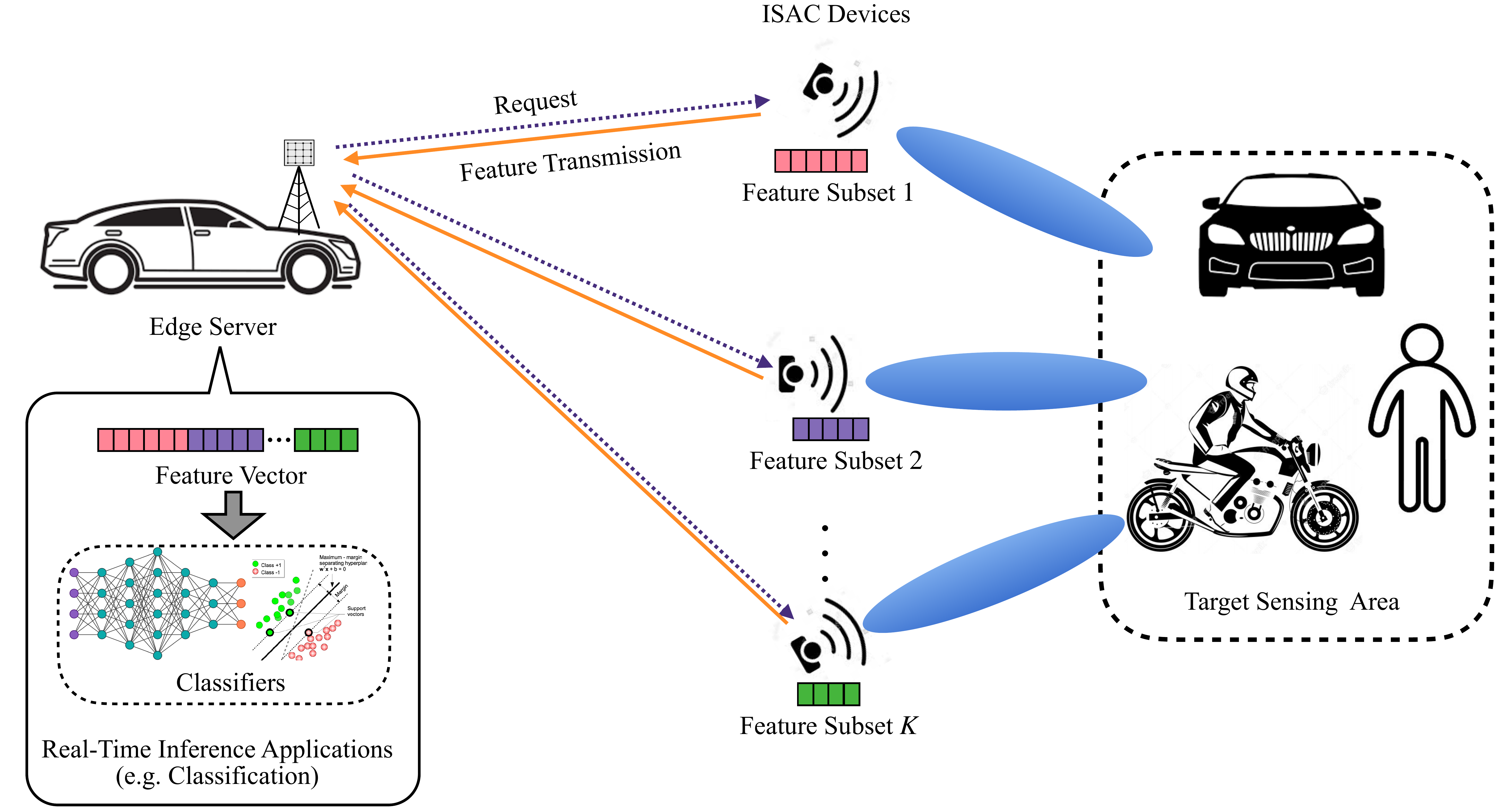}
\caption{Edge inference systems with multi-device sensing.}\label{Fig:SysModel}
\vspace{-1cm}
\end{figure}

\begin{figure}
\centering
\includegraphics[width=0.75\textwidth]{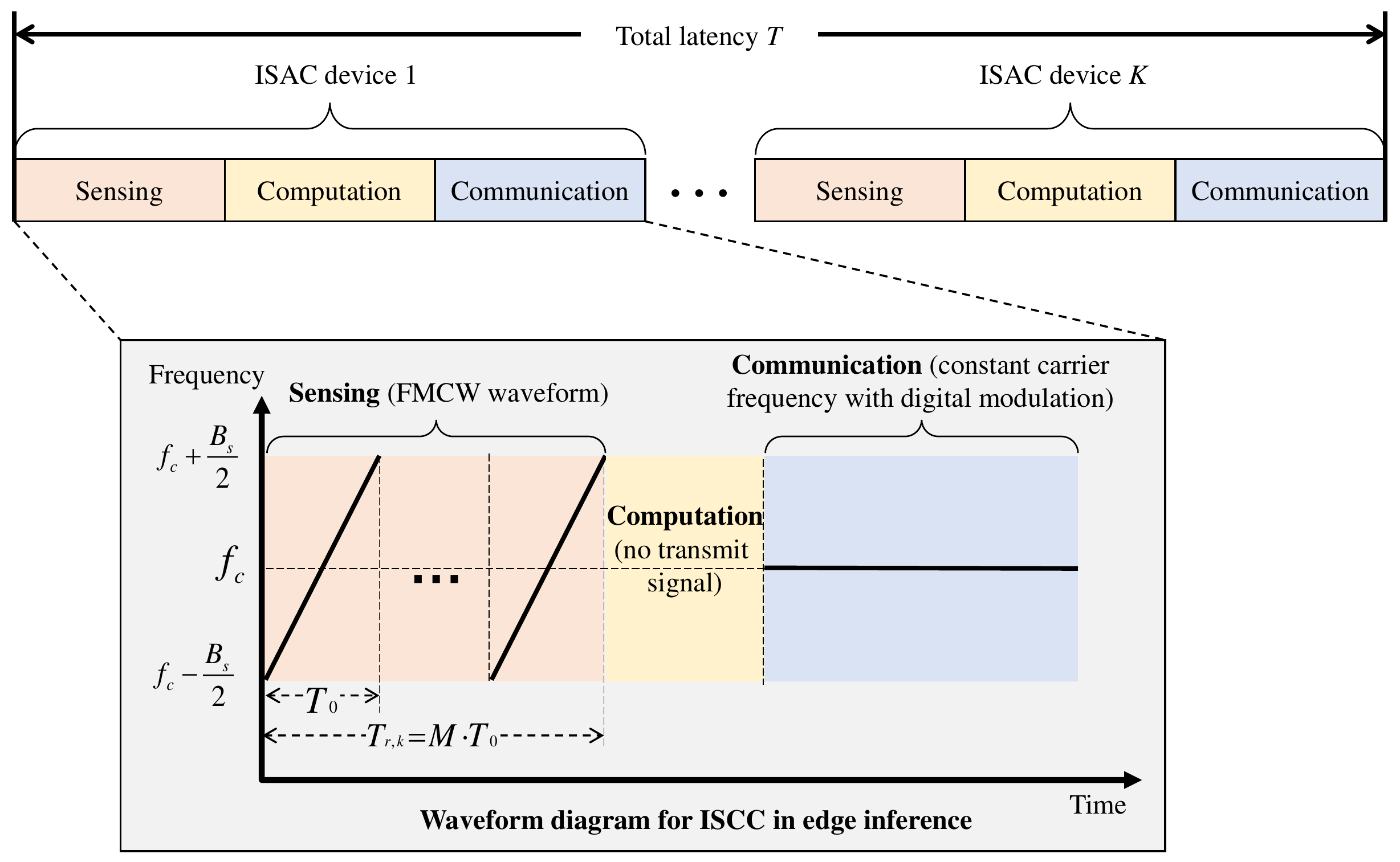}
\caption{Edge inference systems with multi-device sensing.}\label{Fig:TimeDivisionManner}
\vspace{-0.5cm}
\end{figure}

The multi-view radar sensing based edge inference system is shown in Fig. \ref{Fig:SysModel}. There are one mobile edge server  with a single-antenna \emph{access point} (AP) and $K$ single-antenna ISAC devices equipped with DFRC transceivers. In practice, the edge server may correspond to high-mobility vehicles like cars, and the ISAC devices correspond to radar sensors. \emph{Time-division multiple access} (TDMA) is used. The edge server needs to make a real-time decision, such as obstacle detection in the wild, via inferring a well-trained machine learning model. Its features are collected from the ISAC devices. The detailed procedure for data acquisition (sensing), feature extraction and quantization (computation), and feature transmission to the server (communication) at each device is presented in Fig. \ref{Fig:TimeDivisionManner}. Specifically, the server first requests to all devices to sense the environment. Then, the sensing data of each device is processed and quantized locally to a subset of features. Next, all feature subsets are fed back to the edge server via wireless links and are cascaded for completing the reference task. The ISAC devices remain mute to save the energy consumption when there is no request. 

As shown in Fig. \ref{Fig:TimeDivisionManner}, the DFRC transceiver implements ISAC by switching between the sensing mode and communication mode flexibly in a time-division manner using a shared radio-frequency front-end circuit \cite{han2013joint}\footnote{ Practical implementations of the DFRC transceiver via software-defined radio solution have been demonstrated in \cite{han2013joint,ma2021spatial,ma2021frac}. }. In sensing mode,  \emph{frequency-modulated continuous-wave} (FMCW) signal consisting of multiple up-ramp chirps is transmitted \cite{han2013joint}. Then, by processing the received radar echo signals, sensing data  that contain the motion information of the sensing target can be attained at the ISAC devices. In communication mode, constant-frequency carrier modulated by communication data using digital modulation scheme (e.g., QAM) is transmitted.  
The total permitted time to finish the real-time inference task is denoted as $T$. For an arbitrary device, say the $k$-th, its sensing time is denoted as $T_{r,k}$ and its computation time is denoted as $T_{m,k}$, which both are assumed to be constant. 
The communication time to transmit the features is denoted as $T_{c,k}$ and the total communication bandwidth is $B$. The wireless channels are assumed to be static, as the time duration $T$ is short and smaller than the channel coherence time.  The channel gain of the link between the $k$-th device and server is denoted as $H_{c,k}$. The AP is assumed to work as a coordinator and can acquire the \emph{global channel state information} (CSI).

\subsection{Radar Sensing and Feature Generation Model}
In this section, we first model the radar sensing channel for obtaining the sensing data. Then, the signal processing for feature generation is introduced. 

\subsubsection{Sensing Signal}
All ISAC devices transmit linear frequency up-ramp chirp sequences as the sensing signals.  Consider an arbitrary ISAC device, say the $k$-th. A sensing snapshot consists of $M$ chirps, each of which has a duration of $T_{0}=T_{r,k}/M$. The sensing signal in a snapshot is
\begin{align*}
	s_{k}(t)=\sum_{m=0}^{M-1}\text{rect}\left(\frac{t-mT_{0}}{T_{0}}\right) \cdot \cos\left(2 \pi f_{c,k} \left(t-mT_{0}\right) + \pi \mu \left(t-mT_{0}\right)^{2}\right),
\end{align*}
where $\text{rect}(\cdot)$ is the rectangular-shaped pulse function with width of 1 centered at $t = 0$, $f_{c,k}$ is the sensing carrier frequency for the $k$-th ISAC device, $\mu = B_{s}/T_{0}$ is the scope of each chirp, and $B_{s}$ is the bandwidth of the sensing signal. 
The echo signal at time $t$ can be written as
\begin{equation}\label{Eq:SensingModel}
r_k(t) = u_k(t)+\sum_{j=1}^{J}v_{k,j}(t)+n_r(t).
\end{equation}
In \eqref{Eq:SensingModel}, $u_k(t)$ is the desired echo signal directly reflected by the target and is given by
\begin{equation}
u_k(t) = H_{r,k}(t) s_k(t-\tau).
\end{equation}
Here, $H_{r,k}(t)$ is the reflection coefficient including the round-trip path-loss, $\tau$ denotes the round-trip delay, $v_{k,j}(t)$ is the echo signal reflected indirectly by the target from the $j$-th indirect reflection path, which is given by 
\begin{equation}
v_{k,j}(t) = C_{r,k,j}(t)s_k(t - \tau_{j}),
\end{equation}
where $C_{r,k,j}(t)$ and $\tau_{j}$ are the reflection coefficient from and the signal delay of the $j$-th path respectively, $J$ is the total number of indirect reflection paths, and $n_r(t)$ is the Gaussian noise at the sensing receiver. It is assumed that the values of $H_{r,k}(t)$ and $C_{r,k,j}(t)$ can be estimated before sensing.

\subsubsection{Sensing Signal Processing} 
Consider the $k$-th ISAC device, the steps to process the received radar echo signals are as follows:

\textbf{Signal sampling:} For sensing snapshot $m$, the received signal $r_{k}(t)$ in (\ref{Eq:SensingModel}) is sampled into a complex-valued vector $\mathbf{r}_{k,m} \in \mathbb{C}^{MT_{0}f_{s}}$, where $f_{s}$ is the sampling rate. Arrange $\mathbf{r}_{k,m}$ in a two-dimensional data matrix $\mathbf{R}_{k,m} \in \mathbb{C}^{T_{0}f_{s} \times M}$, in which $T_{0}f_{s}$ is the length of the fast-time dimension, and $M$ is the length of the slow-time dimension\footnote{The fast time dimension is referred to as range dimension whose sample intervals can be used for ranging, whereas processing data in the slow-time dimension allows one to estimate the Doppler spectrum at a given fast time dimension.}. 

\textbf{Data filtering:} To mitigate the clutter and extract useful information, we apply a \emph{singular value decomposition} (SVD) based linear filter to $\mathbf{R}_{k,m}$ \cite{Li2021SPAWC}. The data matrix after filtering is given by $\tilde{\mathbf{R}}_{k,m} = \sum_{i=r_{1}}^{r_{2}}\sigma_{i}\mathbf{v}_{i}\mathbf{u}_{i}$, where $\sigma_{i}$, $\mathbf{v}_{i}$, and $\mathbf{u}_{i}$ denote the $i$-th singular value, the $i$-th left-singular vector, and the $i$-th right-singular vector of $\mathbf{R}_{k,m}$, respectively, and $r_{1}$ and $r_{2}$ are empirical parameters. 

\textbf{Feature extraction:} We extract features in the slow-time dimension for inference. First, we transform $\tilde{\mathbf{R}}_{k,m}$ into vector $\tilde{\mathbf{r}}_{k,m} \in \mathbb{C}^{1 \times M}$, i.e., $\tilde{\mathbf{r}}_{k,m} = \mathbf{1}^{T}\tilde{\mathbf{R}}_{k,m}$. Next, PCA is used to extract the principle feature elements from $\tilde{\mathbf{r}}_{k,m}$, and thus make different feature elements uncorrelated. Note that the principle eigen-space can be obtained during the model training process and is obtained at the AP, which is then broadcast to the ISAC devices. The number of extracted feature elements is denoted as $N_k$. Since all the processing steps are linear, the $n_k$-th feature element, following \eqref{Eq:SensingModel}, is given by 
\begin{equation}
	\bar{r}_k(n_k) = \bar{u}_k(n_k)+\sum_{j=1}^{J}\bar{v}_{k,j}(n_k)+\bar{n}_r(n_k),
\end{equation}
where $\bar{u}_k(n_k)$ is the desired ground-truth feature, $\bar{v}_k(n_k)$ is additive information in feature element brought by the clutter signal from the $j$-th path,  $\bar{n}_r(n_k)$ is the noise in feature element.

Each feature element is normalized by the transmit radar sensing power, say $\sqrt{P_{r,k}}$. Specifically, the $n_k$-th feature element is
\begin{equation}\label{Eq:FeatureModel}
\hat{x}(n_k) = \frac{r_k(n)}{ \sqrt{P_{r,k}} } =  x(n_k) + c_{r,k}(n_k) + \dfrac{ n_r(n_k)}{\sqrt{P_{r,k}}},
\end{equation}
where $ x(n_k) = \bar{u}_k(n_k)/\sqrt{P_{r,k}}$ is the ground-true feature and 
\begin{equation}
c_{r,k}(n_k) = \sum\limits_{j=1}^J \dfrac{ \bar{v}_{k,j}(n_k)}{ \sqrt{P_{r,k}} },
\end{equation}
is the normalized clutter. From \eqref{Eq:FeatureModel}, one can observe that the sensed feature is polluted by the clutter, say $c_{r,k}(n_k)$, and the sensing noise $n_r(n_k)$.  According to the \emph{central limit theorem}, $c_{r,k}(n_k)$ is assumed to follow a Gaussian distribution, as the number of independent reflection paths $J$ is large. Its distribution is given as 
\begin{equation}\label{Eq:ClutterDistribution}
c_{r,k}(n_k)\sim \mathcal{N}(0,\sigma_{c,k}^2),
\end{equation}
where $\mathcal{N}(\cdot,\cdot)$ represents the Gaussian distribution and $\sigma_{c,k}^2$ is the constant variance and can be estimated before sensing. The normalized sensing noise also has a Gaussian distribution:
\begin{equation}\label{Eq:SensingNoiseDistribution}
n_r(n_k)/\sqrt{P_{r,k}} \sim \mathcal{N}\left(0,\sigma_r^2/P_{r,k} \right),
\end{equation}
where $\sigma_r^2$ is the noise variance.

The feature subset generated by ISAC device $k$ is 
$\hat{\bf x}_k = \{ \hat{x}(n_k), \; 1\leq n_k \leq N_k  \}$,    
where $N_k$ is the total number of generated feature elements. Furthermore, different feature subsets generated by different ISAC devices are assumed to be independent, as the ISAC devices are sparsely deployed and the corresponding sensing areas are non-overlapping.

\subsection{Quantization Model}
Consider the $k$-th ISAC device, whose feature subset is $\hat{\bf x}_k$. Each feature element is quantized using the same linear quantizer. Specifically, for the $n_k$-th feature element, according to \cite{shlezinger2021deep} and by using high quantization bit range, its quantized version is given by
\begin{equation}\label{Eq:QuantizationModel}
z(n_k) = \sqrt{Q_k} \hat{x}(n_k)+d_k,
\end{equation}
where $\hat{x}(n_k)$ is the original feature element defined in \eqref{Eq:FeatureModel}, $\sqrt{Q_k}$ is the quantization gain, $d_k$ is the approximate Gaussian quantization distortion, given as
\begin{equation}\label{Eq:QunatizationDistortion}
d_k\sim \mathcal{N}(0,\delta_k^2),
\end{equation}
and $\delta_k^2$ is the variance. At the receiver, the quantized feature is recovered as
\begin{equation}\label{Eq:QuantizedFeatureModel}
\tilde{x}(n_k) = \dfrac{z(n_k)}{\sqrt{Q_k}} = \hat{x}(n_k) + \dfrac{d_k}{\sqrt{Q_k}},
\end{equation}
where the notations follow that in \eqref{Eq:QuantizationModel}. Note that in \eqref{Eq:QuantizedFeatureModel}, higher quantization gain, say larger $\sqrt{Q_k}$, can lead to lower quantization distortion in the recovered feature at the receiver. The mutual information of the recovered feature subset $\tilde{\bf x}_k = \{\tilde{x}(1_k), \tilde{x}(2_k),...,\tilde{x}(N_k)\}$ and the generated feature subset $\hat{\bf x}_k$ under the additive Gaussian distortion approximation can be derived as 
\begin{equation}\label{Eq:MutualInformation}
I(\tilde{\bf x}_k;\hat{\bf x}_k) = N_k \log_2\left( 1+\dfrac{Q_k}{\delta_k^2} \right), \; 1\leq k \leq K,
\end{equation}
which is also the overhead of device $k$ for transmitting the feature subset to the server.

\subsection{Discriminant Gain}
Following \cite{lan2021progressive}, we adopt discriminant gain, which is derived from the well-known  KL divergence proposed in \cite{kullback1951information}, as the inference accuracy metric of the classification task.


First, consider an arbitrary feature element generated by the $k$-th ISAC device $\tilde{x}(n_k)$. By substituting $\hat{x}(n_k)$ in \eqref{Eq:FeatureModel} into $\tilde{x}(n_k)$ in \eqref{Eq:QuantizedFeatureModel}, it can be written as
\begin{equation}
\tilde{x}(n_k) = x(n_k) + c_{r,k}(n_k) + \frac{ n_r(n_k)}{\sqrt{P_{r,k}}} + \dfrac{d_k}{\sqrt{Q_k}},
\end{equation}
where the notations follow that in \eqref{Eq:FeatureModel},  \eqref{Eq:ClutterDistribution}, and \eqref{Eq:QuantizedFeatureModel}.

According to \cite{lan2021progressive}, the ground-truth feature element $x(n_k)$ is assumed to have a mixed Gaussian distribution. Its probability density function is 
\begin{equation}\label{Eq:GroundTrueFeatureDistribution}
f\left(x(n_k)\right) = \dfrac{1}{L}\sum\limits_{\ell=1}^L \mathcal{N}\left(\mu_{\ell, n_k}, \sigma_{n_k}^2\right), \; 1\leq n_k \leq N_k, \; 1\leq k \leq K,
\end{equation}
where $L$ is the total number of classes in the inference task, $\mu_{\ell, n_k}$ is the centroid of the $\ell$-th class, and $\sigma_{n_k}^2$ is the variance\footnote{These statistics can be pre-estimated at the AP using the training dataset}. By substituting the distributions of the ground-truth feature in \eqref{Eq:GroundTrueFeatureDistribution}, the clutter distribution in \eqref{Eq:ClutterDistribution}, the normalized sensing noise in \eqref{Eq:SensingNoiseDistribution}, and the quantization distortion in \eqref{Eq:QunatizationDistortion}, into the recovered feature element $\tilde{x}(n_k)$, its distribution can be derived as
\begin{equation}\label{Eq:RecoveredFeatureDistribution}
f\left(\tilde{x}(n_k)\right) = \dfrac{1}{L}\sum\limits_{\ell=1}^L f_{\ell}\left(\tilde{x}(n_k)\right), \; 1\leq n_k \leq N_k, \; 1\leq k \leq K,
\end{equation}
where $f_{\ell}\left(\tilde{x}(n_k)\right)$ is the probability density function of $\tilde{x}(n_k)$ in terms of the $\ell$-th class and is given by
\begin{equation}
f_{\ell}\left(\tilde{x}(n_k)\right)= \mathcal{N}\left(\mu_{\ell, n_k}, \sigma_{n_k}^2 + \sigma_{c,k}^2+\dfrac{\sigma_r^2}{P_{r,k}} + \dfrac{\delta_k^2}{ Q_k }\right), \quad 1\leq \ell \leq L.
\end{equation}

Next, the discriminant gain of $\tilde{x}(n_k)$ can be derived from the well established KL divergence \cite{lan2021progressive}.  Specifically, consider an arbitrary class pair, say classes $\ell$ and $\ell^{'}$. Its discriminant gain is
\begin{equation}\label{Eq:PairDG}
\begin{aligned}
    G_{\ell,\ell^{'}}\left( \tilde{x}(n_k) \right) & = D_{KL} \left[ f_{\ell}\left(\tilde{x}(n_k)\right) \big\| f_{\ell^{'}}\left(\tilde{x}(n_k)\right) \right] + D_{KL} \left[ f_{\ell}\left(\tilde{x}(n_k)\right) \big\| f_{\ell^{'}}\left(\tilde{x}(n_k)\right) \right] \\
    & =  \int_{\tilde{x}(n_k)}f_{\ell}\left(\tilde{x}(n_k)\right) \log\left[ \dfrac{f_{\ell}^{'}\left(\tilde{x}(n_k)\right)}{f_{\ell}\left(\tilde{x}(n_k)\right)} \right] + f_{\ell^{'}}\left(\tilde{x}(n_k)\right) \log\left[ \dfrac{f_{\ell}\left(\tilde{x}(n_k)\right)}{f_{\ell^{'}}\left(\tilde{x}(n_k)\right)} \right]  {\rm d} \tilde{x}(n_k)\\
    & = \dfrac{\left(\mu_{\ell,n_k} - \mu_{\ell^{'},n_k}\right)^2}{\sigma_{n_k}^2 + \sigma_{c,k}^2+ \sigma_r^2/P_{r,k}  + \delta_k^2/Q_k }, \quad \forall (\ell,\ell^{'}),
\end{aligned}
\end{equation}
where $D_{KL} \left[ \cdot \| \cdot \right]$ is the KL divergence defined in \cite{kullback1951information}, and the other notations follow that in \eqref{Eq:RecoveredFeatureDistribution}. It follows that the discriminant gain of the whole feature vector $\tilde{\bf x} = \{\tilde{\bf x}_1,\tilde{\bf x}_2,...,\tilde{\bf x}_K \}$, where $\tilde{\bf x}_k = \{\tilde{x}(1_k), \tilde{x}(2_k),...,\tilde{x}(N_k)\}$, in terms of this class pair is given by
\begin{equation}\label{Eq:PairDG}
\begin{aligned}
 G_{\ell,\ell^{'}}\left( \tilde{\bf x} \right) & = D_{KL} \left[ f_{\ell}\left(\tilde{\bf x}\right) \big\| f_{\ell^{'}}\left(\tilde{\bf x}\right) \right] + D_{KL} \left[ f_{\ell}\left(\tilde{\bf x}\right) \big\| f_{\ell^{'}}\left(\tilde{\bf x}\right) \right] = \sum\limits_{k=1}^K \sum\limits_{n_k=1}^{N_k} G_{\ell,\ell^{'}}\left( \tilde{x}(n_k) \right),
\end{aligned}
\end{equation}
since different feature elements in $\tilde{\bf x}$ are independent. The overall discriminant gain of $\tilde{\bf x}$ is defined as the average of all class pairs:
\begin{equation}\label{Eq:DiscriminantGainSystem}
G =    \dfrac{2}{L(L-1)}\sum\limits_{n_k=1}^{N_k}\sum\limits_{\ell^{'}=1}^L \sum\limits_{\ell<\ell^{'}} G_{\ell,\ell^{'}}\left( \tilde{x}(n_k) \right).
\end{equation}

\section{Problem Formulation \& Simplification}

\subsection{Problem Formulation}
Our objective is to maximize the total discriminant gain in \eqref{Eq:DiscriminantGainSystem} under the constraints on latency, successful transmission, and energy. By substituting \eqref{Eq:PairDG} into \eqref{Eq:DiscriminantGainSystem}, the objective can be written as
\begin{equation}\label{Eq:Objective}
\max\limits_{P_{c,k},P_{r,k},T_{c,k},Q_k}\;\; G =  \dfrac{2}{L(L-1)}\sum\limits_{k=1}^K \sum\limits_{n_k=1}^{N_k} \sum\limits_{\ell^{'}=1}^L \sum\limits_{\ell<\ell^{'}}\dfrac{\left(\mu_{\ell,n_k} - \mu_{\ell^{'},n_k}\right)^2}{\sigma_{n_k}^2 + \sigma_{c,k}^2+  \sigma_r^2/P_{r,k}  + \delta_k^2/Q_k},
\end{equation}
where the notations follow that in \eqref{Eq:PairDG}. Next, we formulate the various constraints.

\subsubsection{Latency Constraint} 
The total allocated sensing, computation, and communication time should be less than the permitted latency of the real-time inference task:
\begin{equation}
\text{(C1)}\quad \sum\limits_{k=1}^{K} ( T_{r,k}+ T_{m,k} + T_{c,k} ) \leq T,
\end{equation}
where $T_{r,k}$, $T_{m,k}$, and $T_{c,k}$ are the  constant sensing time, the constant computation time, 
and  the allocated communication time of ISAC device $k$ respectively,  and $T$ is the permitted latency to finish the task.

\subsubsection{Successful Transmission Constraint}
To ensure successful transmission of the quantized feature subset to the receiver, the mutual information between the generated feature subset $\hat {\bf x}_k$ and the recovered one $\tilde {\bf x}_k$ should be less than the channel capacity as formally stated below \cite{Park2013TSP-quantization}:
\begin{equation}\label{Eq:TransmitCondition1}
I(\tilde{\bf x}_k;\hat{\bf x}_k) \leq R_k, \; 1\leq k \leq K,
\end{equation}
where $R_k$ is the channel capacity of ISAC device $k$. 
It is given by
\begin{equation}\label{Eq:ChannelCapacity}
R_k = T_{c,k} B \log_2\left(1 + \dfrac{P_{c,k} H_{c,k}}{ \delta_c^2 } \right),\; 1\leq k \leq K,
\end{equation}
where $B$ is the system bandwidth, $\delta_c^2$ is the channel noise power, $T_{c,k}$ is the allocated time slot, $P_{c,k}$ is the transmit power, and $H_{c,k}$ is the channel gain. By substituting the mutual information in  \eqref{Eq:MutualInformation} and the data rate in \eqref{Eq:ChannelCapacity} into the transmission constraint in \eqref{Eq:TransmitCondition1}, it can be written as
\begin{equation}\label{Eq:TransmitCondition2}
\text{(C2)}\quad N_k \log_2\left( 1+\dfrac{Q_k}{\delta_k^2} \right) \leq T_{c,k} B \log_2\left(1 + \dfrac{P_{c,k} H_{c,k}}{ \delta_c^2 } \right),\; 1\leq k \leq K.
\end{equation}

\subsubsection{Energy Constraint}
The energy consumption of each ISAC device should be bounded:
\begin{equation}
\text{(C3)}\quad P_{r,k}T_{r,k} + E_{m,k} + P_{c,k}T_{c,k}\leq E_k,\;\; 1\leq k \leq K,
\end{equation}
where $P_{r,k}$, $P_{c,k}$, $T_{r,k} $,  $T_{c,k} $, $E_{m,k}$, and $E_k$ are the sensing power, the transmit power, the constant sensing time, the communication time, the constant computation energy consumption, 
and the energy threshold of ISAC device $k$, respectively.

Under the three kinds of constraints above, the problem of maximizing discriminant gain is formulated as 
\begin{equation}\text{(P1)}\quad
\begin{aligned}
\max\limits_{P_{c,k},P_{r,k},T_{c,k},Q_k}\;\; &G =  \dfrac{2}{L(L-1)}\sum\limits_{k=1}^K \sum\limits_{n_k=1}^{N_k} \sum\limits_{\ell^{'}=1}^L \sum\limits_{\ell<\ell^{'}}\dfrac{\left(\mu_{\ell,n_k} - \mu_{\ell^{'},n_k}\right)^2}{\sigma_{n_k}^2 + \sigma_{c,k}^2+  \sigma_r^2/P_{r,k}  + \delta_k^2/Q_k},\\
\text{s.t.}\;\; &P_{c,k},P_{r,k},T_{c,k},Q_k \in \mathbb{R}^+ ,\quad 1 \leq k \leq K,\\
& \text{(C1)} \sim \text{(C3)}.
\end{aligned}
\end{equation}
(P1) is a non-convex problem due to the non-convexity of the objective function and Constraints (C2) and (C3) therein.   Although the discriminant gain maximization problem is investigated in \cite{lan2021progressive} via progress feature transmission, this work is the first to enhance the inference performance from a systematic view, i.e., the integration of sensing, computation and communication.  In the sequel, an equivalent simplified problem is derived.

\subsection{Problem Simplification}
To simplify (P1), the following variable transformations are applied:
\begin{equation}\label{Eq:VariablesTransform}
  S_k = \dfrac{\sigma_r^2}{P_{r,k}},\quad  D_k = \dfrac{\delta_k^2}{Q_k}, \quad E_{c,k} = P_{c,k}T_{c,k}, 
\end{equation}
where $S_k$, $D_k$, and $E_{c,k}$ can be interpreted as the normalized sensing noise power, the normalized quantization distortion, and the communication energy consumption of ISAC device $k$, respectively. By substituting \eqref{Eq:VariablesTransform} into (P1), it can be equivalently derived as
\begin{equation}\text{(P2)}\quad
\begin{aligned}
\max\limits_{E_{c,k},S_k,T_{c,k},D_k}\;\; &G = \dfrac{2}{L(L-1)}\sum\limits_{k=1}^K \sum\limits_{n_k=1}^{N_k} \sum\limits_{\ell^{'}=1}^L \sum\limits_{\ell<\ell^{'}}\dfrac{\left(\mu_{\ell,n_k} - \mu_{\ell^{'},n_k}\right)^2}{\sigma_{n_k}^2 + \sigma_{c,k}^2+ S_k  + D_k },\\
\text{s.t.}\;\; & P_{c,k},P_{r,k},T_{c,k},Q_k \in \mathbb{R}^+ ,\quad 1 \leq k \leq K,\\
& \sum\limits_{k=1}^{K} ( T_{r,k} + T_{m,k}+ T_{c,k} ) \leq T, \\
& N_k \log_2\left( 1+\dfrac{1}{D_k} \right) \leq T_{c,k} B \log_2\left(1 + \dfrac{E_{c,k} H_{c,k}}{ T_{c,k} \delta_c^2 } \right),\; 1\leq k \leq K, \\
& \dfrac{\sigma_r^2T_{r,k}}{S_k}+ E_{m,k} + E_{c,k}\leq E_k,\;\; 1\leq k \leq K.
\end{aligned}
\end{equation}
In (P2), all constraints are convex but the objective function (in the form of summation over multiple ratios) to be maximized is non-concave, thus making (P2) non-convex. To tackle the problem, the sum-of-ratios method is used in the following.

\section{Optimal ISCC Scheme}
\begin{figure}
	\centering
	\includegraphics[width=0.8\textwidth]{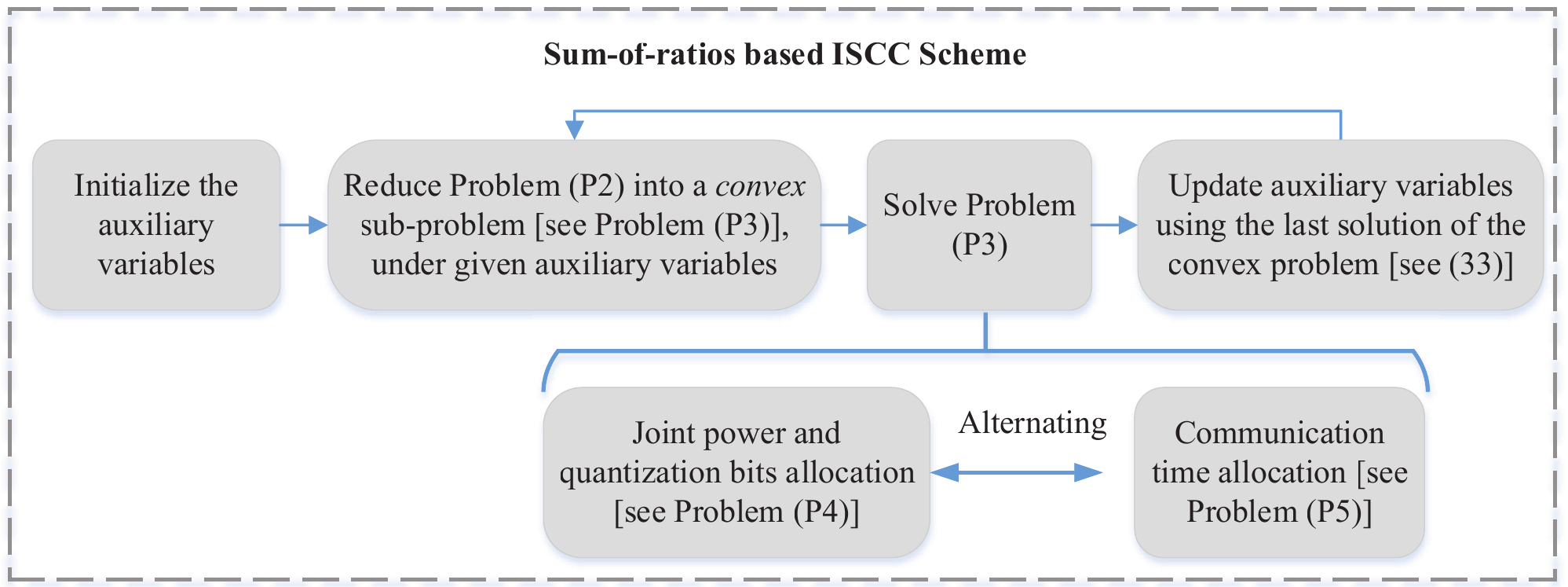}
	\caption{Solution Methodology of the ISCC scheme.}
	\vspace{-0.5cm}
	\label{fig:solutionprocess}
\end{figure}

In this section, an optimal ISCC scheme for joint sensing \& transmit power, time, and quantization bits allocation, is proposed to solve (P2). The solution process is presented in Fig. \ref{fig:solutionprocess}. Specifically, (P2) is optimally tackled by an iterative method, called \emph{sum-of-ratios}. In each iteration,  the auxiliary variables are first introduced to derive a convex problem from (P2), called \emph{sum of weighted  distortion minimization}. Then, the convex problem is addressed by alternately solving the problem of joint power and quantization bits allocation and the problem of communication time allocation.

\subsection{The Sum-of-Ratios Method}
In this part, the sum-of ratios method in \cite{jong2012efficient} is utilized to optimally address (P2) by alternating between two steps: 1) solving a convex sub-problem, that is derived from (P2) to minimize the sum of weighted sensing and quantization distortion under given discriminant gains, and 2) updating the discriminant gains using the solved distortion level of sensing and quantization. These two steps iterate till convergence. The detailed procedure is elaborated in the sequel.

To begin with, we show that the sum-of ratios method can be applied to solve (P2), as shown in the lemma below.
\begin{lemma}\label{Lma:SumOfRatios}
The objective function of (P2) is the sum of multiple quasi-linear ratios. (P2) can be optimally solved using the sum-of-ratios method.
\end{lemma}
\proof See Appendix \ref{Apdx:LmaSumOfRatios}.

Based on Lemma \ref{Lma:SumOfRatios}, the detailed solution process via using the sum-of-ratios method is presented as follows. First, the objective function of (P2) is rewritten as 
\begin{equation}\label{Eq:DG11}
G = \sum\limits_{k=1}^K \sum\limits_{n_k=1}^{N_k} \sum\limits_{\ell^{'}=1}^L \sum\limits_{\ell<\ell^{'}} \dfrac{ \mathcal{A}_{\ell,\ell^{'},n_k}  }{ \mathcal{B}_{\ell,\ell^{'},n_k}\left( S_k, D_k\right) },
\end{equation}
where $ \mathcal{A}_{\ell,\ell^{'},n_k} $ and $\mathcal{B}_{\ell,\ell^{'},n_k}\left( S_k, D_k\right)$ are
\begin{equation}\label{Eq:Fraction}
\left\{
\begin{aligned}
& \mathcal{A}_{\ell,\ell^{'},n_k}  = 1, \\
& \mathcal{B}_{\ell,\ell^{'},n_k}\left( S_k, D_k\right) = \dfrac{L(L-1)(\sigma_{n_k}^2 + \sigma_{c,k}^2+ S_k  + D_k) }{2 \left(\mu_{\ell,n_k} - \mu_{\ell^{'},n_k}\right)^2},
\end{aligned}
\right. \quad \forall (\ell,\ell^{'},n_k).
\end{equation}
We then creat the following sub-problem:
\begin{equation}\text{(P3)}\quad
\begin{aligned}
\max_{E_{c,k},S_k,T_{c,k},D_k}\;\;\;\; &\sum\limits_{k=1}^K \sum\limits_{n_k=1}^{N_k} \sum\limits_{\ell^{'}=1}^L \sum\limits_{\ell<\ell^{'}}x_{\ell,\ell^{'},n_k} \left[ \mathcal{A}_{\ell,\ell^{'},n_k}  - y_{\ell,\ell^{'},n_k}  \mathcal{B}_{\ell,\ell^{'},n_k} \left(S_k,D_k\right) \right],\\
\text{s.t.}\;\; & \text{All constraints in (P2)}, 
\end{aligned}
\end{equation}
where $\left\{  x_{\ell,\ell^{'},n_k} \right\}$ and $\left\{  y_{\ell,\ell^{'},n_k} \right\}$ are the introduced auxiliary variables, and $\mathcal{A}_{\ell,\ell^{'},n_k}$ and $ \mathcal{B}_{\ell,\ell^{'},n_k} \left(S_k,D_k\right)$ are defined in \eqref{Eq:Fraction}. In (P3), each term in the objective function is a scale of the sum of sensing noise power $\{S_k\}$ and quantization distortion $\{D_k\}$, giving its name of \emph{sum of weighted distortion minimization problem}. It is easy to show that (P3) is convex.


Next, according to \cite{jong2012efficient} and Theorem 1 in \cite{yu2015joint}, (P2) can be optimally addressed by alternating between optimally solving the sub-problem in (P3) under given auxiliary variables $\left\{  x_{\ell,\ell^{'},n_k} \right\}$ and $\left\{  y_{\ell,\ell^{'},n_k} \right\}$, and updating them based on the correspondingly obtained solution. Hence, based on the convexity of (P3), (P2) can be optimally solved by iteratively performing the following two steps till convergence.
\begin{itemize}
\item \emph{Step 1}: Optimally solving (P3) with given auxiliary variables $\left\{  x_{\ell,\ell^{'},n_k} \right\}$ and $\left\{  y_{\ell,\ell^{'},n_k} \right\}$. 

\item \emph{Step 2}: Updating the auxiliary variables $\left\{  x_{\ell,\ell^{'},n_k} \right\}$  and $\left\{ y_{\ell,\ell^{'},n_k} \right\}$  as
\begin{equation}
\begin{aligned}
& x_{\ell,\ell^{'},n_k}  = \dfrac{1}{\mathcal{B}_{\ell,\ell^{'},n_k} \left(S_k,D_k\right)},\;\forall (\ell,\ell^{'},n_k), \\
& y_{\ell,\ell^{'},n_k}  = \dfrac{\mathcal{A}_{\ell,\ell^{'},n_k}}{\mathcal{B}_{\ell,\ell^{'},n_k} \left(S_k,D_k\right)} = \dfrac{1}{\mathcal{B}_{\ell,\ell^{'},n_k} \left(S_k,D_k\right)}, \; \forall (\ell,\ell^{'},n_k),
\end{aligned}
\end{equation}
where $ \mathcal{B}_{\ell,\ell^{'},n_k} \left(S_k,D_k\right)$ is defined in \eqref{Eq:Fraction}. From the above equation, it can be observed that 
\begin{equation}\label{Eq:EqualVar}
 x_{\ell,\ell^{'},n_k} = y_{\ell,\ell^{'},n_k},
\end{equation}
and they are the discriminant gain of feature $n_k$ between the classes $\ell$ and $\ell^{'}$.
\end{itemize}
The above process can be interpreted as iterating over addressing the sum of weighted distortion minimization problem under given discriminant gain, and updating the discriminant gain using the solved sensing and communication distortion level. 

\subsection{An Alternating Method for Solving (P3)}
In this section, an alternating algorithm is proposed to solve the convex sub-problem in (P3) with given auxiliary variables $\left\{  x_{\ell,\ell^{'},n_k} \right\}$  and $\left\{ y_{\ell,\ell^{'},n_k} \right\}$. This allows closed-form solutions with structural properties and can achieve low computational complexity. 
Next, the two sub-problems are first introduced, followed by a summary of the alternating algorithm.

\subsubsection{Joint Power and Quantization Bits Allocation}
In this case, the communication time, say $\{T_{c,k}\}$, is given. By substituting \eqref{Eq:EqualVar} and   $\mathcal{A}_{\ell,\ell^{'},n_k}$ in \eqref{Eq:Fraction}, (P3) can be written as
\begin{equation}\text{(P4)}\;
\begin{aligned}
\max_{E_{c,k},S_k,D_k}\;\;\;\; &\sum\limits_{k=1}^K \sum\limits_{n_k=1}^{N_k} \sum\limits_{\ell^{'}=1}^L \sum\limits_{\ell<\ell^{'}} \left[ y_{\ell,\ell^{'},n_k}   - y_{\ell,\ell^{'},n_k}^2  \mathcal{B}_{\ell,\ell^{'},n_k} \left(S_k,D_k\right) \right],\\
\text{s.t.}\;\; & E_{c,k},S_k,D_k \in \mathbb{R}^+ ,\quad 1 \leq k \leq K,\\
&N_k\log_2\left( 1+\dfrac{1}{D_k} \right)\leq T_{c,k} B \log_2\left( 1+  \dfrac{ E_{c,k} H_{c,k} }{ T_{c,k} \delta_c^2 } \right),\;\; 1\leq k \leq K,\\
&\dfrac{\sigma_r^2T_{r,k}}{S_k} + E_{m,k}+ E_{c,k}  \leq E_k,\;\; 1\leq k \leq K,\\
\end{aligned}
\end{equation}
which is a convex problem.
The \emph{Karush-Kuhn-Tucker} (KKT) conditions are used to solve (P4). The Lagrangian is given by
\begin{equation}
\begin{aligned}
\mathcal{L}_{\text{P4}}  = &- \sum\limits_{k=1}^K \sum\limits_{n_k=1}^{N_k} \sum\limits_{\ell^{'}=1}^L \sum\limits_{\ell<\ell^{'}}    \left[ y_{\ell,\ell^{'},n_k}   - y_{\ell,\ell^{'},n_k}^2  \mathcal{B}_{\ell,\ell^{'},n_k} \left(S_k,D_k\right) \right],\\
&+\sum\limits_{k=1}^K\alpha_k \left[ N_k\log_2\left( 1+\dfrac{1}{D_k} \right)- T_{c,k} B \log_2\left( 1+  \dfrac{ E_{c,k} H_{c,k} }{ T_{c,k} \delta_c^2 } \right)  \right],\\
&+\sum\limits_{k=1}^K \beta_k \left( \dfrac{T_{r,k}}{S_k} + E_{m,k}+ E_{c,k}- E_k \right),
\end{aligned}
\end{equation}
where $\{\alpha_k\geq 0\}$ and $\{\beta_k\geq 0\}$ are the corresponding Lagrange multipliers,  and $\mathcal{B}_{\ell,\ell^{'},n_k} \left(S_k,D_k\right)$ is defined in \eqref{Eq:Fraction}.

The first KKT condition can be written as
\begin{equation}
\dfrac{\partial \mathcal{L}_{\text{P4}} }{\partial S_k } = \sum\limits_{\ell^{'}=1}^L \sum\limits_{\ell<\ell^{'}}\left(  y_{\ell,\ell^{'},n_k}^2 \times \dfrac{\partial \mathcal{B}_{\ell,\ell^{'},n_k} }{\partial S_k }  \right)-  \dfrac{\beta_k T_{r,k}}{S_k^2} = 0,\quad 1\leq k \leq K,
\end{equation}
where, according to \eqref{Eq:Fraction}, 
\begin{equation}
 \dfrac{\partial B_{\ell,\ell^{'},n_k} }{ \partial S_k } = \dfrac{L(L-1) }{2 \left(\mu_{\ell,n_k} - \mu_{\ell^{'},n_k}\right)^2}.
\end{equation}
It follows that 
\begin{equation}\label{Eq:SensingNoiseControl}
\dfrac{1}{S_k} = \sqrt{   \sum\limits_{\ell^{'}=1}^L \sum\limits_{\ell<\ell^{'}}    \dfrac{L(L-1) y_{\ell,\ell^{'},n_k}^2 }{2 \left(\mu_{\ell,n_k} - \mu_{\ell^{'},n_k}\right)^2}  \times \dfrac{1}{ \beta_k T_{r,k} } }.
\end{equation}
By substituting $S_k$ in \eqref{Eq:VariablesTransform} into \eqref{Eq:SensingNoiseControl}, the following optimal sensing power allocation scheme can be obtained. 
\begin{lemma}
The optimal sensing power for ISAC device $k$ must satisfy
\begin{equation}\label{Eq:SensingPowerAllocation}
P_{r,k} = \sigma_r^2 \times \sqrt{   \sum\limits_{\ell^{'}=1}^L \sum\limits_{\ell<\ell^{'}}    \dfrac{L(L-1) y_{\ell,\ell^{'},n_k}^2 }{2 \left(\mu_{\ell,n_k} - \mu_{\ell^{'},n_k}\right)^2}  \times \dfrac{1}{ \beta_k T_{r,k} } },\quad 1\leq k \leq K,
\end{equation}
where $\{\beta_k\}$ are the Lagrangian multipliers.
\end{lemma}
From \eqref{Eq:SensingPowerAllocation}, we conclude the following. Consider an arbitrary ISAC device, say the $k$-th one. First, if the number of classes $L$ is large, or the required discriminant gains $\{y_{\ell,\ell^{'},n_k}\}$ are large, more power should be allocated for sensing. Then, if the centroid distances, say $\{\left(\mu_{\ell,n_k} - \mu_{\ell^{'},n_k}\right)^2\}$, are large, or the sensing noise variance $\sigma_r^2$ is small, the required sensing power can be reduced. In addition, long sensing time, i.e., larger $T_{r,k}$, can also reduce the required sensing power.

The second KKT condition is given by 
\begin{equation}
\dfrac{\partial \mathcal{L}_{\text{P4}} }{\partial D_k } = \sum\limits_{\ell^{'}=1}^L \sum\limits_{\ell<\ell^{'}}\left( y_{\ell,\ell^{'},n_k}^2\times \dfrac{\partial \mathcal{B}_{\ell,\ell^{'},n_k} }{\partial D_k }  \right)-  \dfrac{\alpha_k N_k\ln2 }{D_k(D_k+1)} = 0,\quad 1\leq k \leq K,
\end{equation}
which, by substituting  $\mathcal{B}_{\ell,\ell^{'},n_k} \left(S_k,D_k\right)$ in \eqref{Eq:Fraction}, can be derived as
\begin{equation}\label{Eq:QuantizationDistortionControl}
D_k = \sqrt{ \dfrac{1}{4} +  \dfrac{ \alpha_k N_k \ln 2}{    \sum\limits_{\ell^{'}=1}^L \sum\limits_{\ell<\ell^{'}}    \dfrac{L(L-1) y_{\ell,\ell^{'},n_k}^2 }{2 \left(\mu_{\ell,n_k} - \mu_{\ell^{'},n_k}\right)^2} } } - \dfrac{1}{2},\quad 1\leq k\leq K,
\end{equation}
where $\{\alpha_k\}$ are the Lagrangian multipliers.
By substituting $D_k$ in \eqref{Eq:VariablesTransform} into \eqref{Eq:QuantizationDistortionControl}, we obtain the following lemma.
\begin{lemma}
The optimal quantization gain satisfies
\begin{equation}\label{Eq:QuantiaztionBitsAllocation}
Q_k = \dfrac{\delta_k^2}{D_k}, \quad 1\leq k \leq K,
\end{equation}
where $\delta_k^2$ is the quantization distortion and $D_k$ is defined in \eqref{Eq:QuantizationDistortionControl}.
\end{lemma}
Several observations can be made from \eqref{Eq:QuantiaztionBitsAllocation}. For an arbitrary ISAC device, say the $k$-th, larger number of classes $L$, larger number of feature elements $N_k$, and larger required discriminant gains $\{ y_{\ell,\ell^{'},n_k}\}$, call for greater quantization gain (or level), as it requires more fine-grained feature representations to increase the differentiability among them. In addition, larger centroid distances between classes, say $\{\left(\mu_{\ell,n_k} - \mu_{\ell^{'},n_k}\right)^2\}$, require smaller quantization gain, since different classes are well separated and thus low-resolution feature representation is fine for discriminating them.

The third KKT condition can be written as
\begin{equation}
\dfrac{\partial \mathcal{L}_{\text{P4}} }{\partial E_{c,k} } =  -\dfrac{ \alpha_k B T_{c,k} H_{c,k} }{ (E_{c,k}H_{c,k} + T_{c,k}\delta_c^2 )\ln2}  + \beta_k = 0.
\end{equation}
It follows that 
\begin{equation}\label{Eq:EnergyAllocation}
	E_{c,k} = \max\left\{\dfrac{ \alpha_k B T_{c,k}}{\beta_k \ln2} - \dfrac{T_{c,k}\delta_c^2}{H_{c,k}},\quad 0\right\}.
\end{equation}
By substituting $E_{c,k}$ in \eqref{Eq:VariablesTransform} into \eqref{Eq:EnergyAllocation}, we have the following optimal power allocation.
\begin{lemma}
The optimal communication power for each ISAC device should be
\begin{equation}\label{Eq:CommunicationPowerAllocation}
P_{c,k} = \max\left\{\dfrac{\alpha_k B }{ \beta_k \ln2  } - \dfrac{ \delta_c^2 }{ H_{c,k} },\quad 0\right\},\quad 1\leq k \leq K.
\end{equation} 
\end{lemma}

Based on the results above, the primal-dual method can be used to solve (P4), as summarized in Algorithm \ref{Alg:PowerQuantizationAllocation}.
\begin{algorithm}
\caption{Joint Power and Quantization Bits Allocation}\label{Alg:PowerQuantizationAllocation}

1: {\bf Input:} Channel gains $\{H_{c,k}\}$, auxiliary variables $\{y_{\ell,\ell^{'},n_k}\}$, feature elements' class centroids $\{\mu_{\ell,n_k}\}$ and variances $\{\sigma_{n_k}^2\}$, and the given communication latencies $\{T_{c,k}\}$.
  
2: {\bf Initialize} $\{\alpha_k^{(0)}\}$, $\{\beta_k^{(0)}\}$, the step sizes $\{\eta_{\alpha_k} \}$ and $\{\eta_{\beta_k} \}$, and $i=0$.

3: {\bf Loop}

4: \quad Solve $\{S_k\}$, $\{D_k\}$, and $\{E_{c,k}\}$ using \eqref{Eq:SensingNoiseControl}, \eqref{Eq:QuantizationDistortionControl}, and \eqref{Eq:EnergyAllocation}, respectively.

5: \quad Update the multipliers as 
\begin{equation*}
\left\{
\begin{aligned}
&\alpha_k^{(i+1)} = \max\left\{\alpha_k^{(i)} +\eta_{\alpha_k} \dfrac{ \partial \mathcal{L}_{\text{P4} }}{\partial \alpha_k}, \quad 0\right\}, \;1\leq k \leq K, \\
&\beta_k^{(i+1)} = \max\left\{\beta_k^{(i)} +\eta_{\beta_k} \dfrac{ \partial \mathcal{L}_{\text{P4} }}{\partial \beta_k}, \quad 0\right\}, \;1\leq k \leq K,
\end{aligned}
\right.
\end{equation*}

6: \quad $i=i+1$.

7: {\bf Until Convergence}

8: Calculate $\mathcal{B}_{\ell,\ell^{'},n_k} \left(S_k,D_k\right)$ using \eqref{Eq:Fraction}.

9: {\bf Output:} $\left\{\mathcal{B}_{\ell,\ell^{'},n_k} \left(S_k,D_k\right)\right\}$,  $\{S_k\}$, $\{D_k\}$, and $\{E_{c,k}\}$.

\end{algorithm}

\subsubsection{Communication Time Allocation}
In this case, the normalized sensing noise power $\{S_k\}$, communication energy $\{E_{c,k}\}$, and normalized quantization distortion $\{D_k\}$ are first solved by Algorithm \ref{Alg:PowerQuantizationAllocation}. To determine the communication time allocation $\{T_{c,k}\}$, a feasibility problem of (P3) is first derived, as shown in (P5). It obtains the minimum required time, dented as $T^*$, under given weighted distortion determined by $\{S_k\}$, $\{E_{c,k}\}$, and $\{D_k\}$. Then, following the methods used in \cite{wen2020joint} and  \cite{wen2021adaptive}, the tractability of (P3) under the current weighted distortion is determined by the comparison between $T^*$ and the permitted latency $T$, as described below. 
\begin{itemize}
\item \emph{Case of $T^*>T$:} In this case, the given $\{S_k\}$, $\{D_k\}$, and $\{E_{c,k}\}$ are not in the feasible region of (P3).  The reason is that the latency constraint therein cannot be satisfied. To this end, the latency of all ISAC devices should be reduced to satisfy the constraint \footnote{If the initial point is feasible, the solution will not fall into this case by using the sequel algorithm.}.

\item  \emph{Case of $T^*<T$:} In this case,  more time can be allocated to all ISAC devices to achieve discriminant gain in (P3).

\item \emph{Case of $T^*=T$:} The current time allocation is optimal.
\end{itemize}
Based on the observations above, for the first two cases, a time updating rule is proposed to re-allocate the remaining (exceeding) time $(T-T^*)$ to all devices, which can guarantee (P3) is feasible in the next iterations, and  reduce the total weighted distortion. In the sequel, the detailed procedure is described.

First, the feasibility problem is given by 
\begin{equation}\text{(P5)}\quad 
\begin{aligned}
T^* = \min_{T_{c,k}}\;\; & \sum\limits_{k=1}^{K} (T_{c,k} + T_{m,k}+ T_{r,k}) \\
\text{s.t.}\;\;& T_{c,k} \in \mathbb{R}^+, \quad 1\leq k \leq K,\\
&N_k\log_2\left( 1+\dfrac{1}{D_k} \right)\leq T_{c,k} B\log_2\left( 1+  \dfrac{ E_{c,k} H_{c,k} }{ T_{c,k} \delta_c^2 } \right),\; 1\leq k \leq K.\\
\end{aligned}
\end{equation}
To solve (P5), its Lagrange function is derived as 
\begin{equation}
\mathcal{L}_{\text{P5}} = \sum\limits_{k=1}^{K} (T_{c,k} + T_{m,k}+ T_{r,k}) + \sum\limits_{k=1}^{K} \lambda_k  \left[ N_k\log_2\left( 1+\dfrac{1}{D_k} \right)- T_{c,k} B\log_2\left( 1+  \dfrac{ E_{c,k} H_{c,k} }{ T_{c,k} \delta_c^2 } \right) \right],
\end{equation}
where $\{\lambda_k\geq 0\}$ are the Lagrangian multipliers. As (P5) is convex, the primal-dual method can be used to obtain the optimal solution, where the optimizer are denoted as $\{T_{c,k}^*\}$.  

Then, the communication time updating to re-allocate the remaining (exceeding) time $(T-T^*)$ is designed as follows:
\begin{equation}\label{Eq:CommunicationTimeUpdate}
T_{c,k} = T_{c,k}^* +  \dfrac{\gamma_k}{\sum\nolimits_{k=1}^K \gamma_k}\times (T - T^*),\quad 1\leq k \leq K,
\end{equation}
where $T^*$ is the obtained optimal total duration, $T_{c,k}^*$ is the solved optimal communication time of ISAC device $k$, $\gamma_k$ is defined as
\begin{equation}
\gamma_k =   \dfrac{\partial \mathcal{L}_{\text{P5}} } {\partial \lambda_k} \bigg|_{\lambda_k =\lambda_k^*} =  N_k\log_2\left( 1+\dfrac{1}{D_k} \right) - T_{c,k} B\log_2\left( 1+  \dfrac{ E_{c,k} H_{c,k} }{ T_{c,k}^* \delta_c^2 } \right).
\end{equation}
Several observations can be made from \eqref{Eq:CommunicationTimeUpdate}. First, if the current total weighted distortion is not feasible in the given delay, i.e., $T^*>T$, using the updating rule in \eqref{Eq:CommunicationTimeUpdate} can make (P3) feasible in the next iterations.  Then, it is observed $\gamma_k$ represents the \emph{ throughput gap} of device $k$ between the required communication load for reliably transmitting the quantized feature subset and the available channel capacity. If the minimum required latency is less than the permitted one, i.e., $T^*<T$, the updating rule indicates that the device requiring more communication capacity is allocated with more time.

\begin{proposition}[Enhanced Discriminant Gain via Additional Time Allocation]\label{Prop:MonotonicTimeAllocation}
The time updating rule in \eqref{Eq:CommunicationTimeUpdate} leads to smaller weighted distortion level for (P3) and results in enhanced discriminant gain.
\end{proposition}

\proof See Appendix \ref{Apdx:LmaMonotonicTimeAllocation}.

Overall, the primal dual method to solve (P5) and the communication time updating are summarized in Algorithm \ref{Alg:CommunicationTimeAllocation}, where $\eta_{\lambda_k} $ and $\eta_k$ are the step sizes, and 
\begin{equation}
\dfrac{ \partial \mathcal{L}_{\text{P5} }}{\partial T_{c,k}} = 1-\lambda_k  \left[ B\log_2\left( 1+  \dfrac{ E_{c,k} H_{c,k} }{ T_{c,k} \delta_c^2 } \right) + \dfrac{E_{c,k} H_{c,k}   }{ (E_{c,k} H_{c,k} + T_{c,k} \delta_c^2  )\ln2}\right], \;1\leq k \leq K,
\end{equation}
where the notations follow those in \eqref{Eq:ChannelCapacity} and \eqref{Eq:VariablesTransform}.

\begin{algorithm}[!th]
\caption{Communication Time Allocation for solving (P5)}\label{Alg:CommunicationTimeAllocation}

1:  {\bf Input:}  $\{S_k\}$, $\{E_{c,k}\}$, and $\{D_k\}$.
  
2:  {\bf Initialize} $\{\lambda_k^{(0)}\}$, the step sizes $\{\eta_{\lambda_k}\}$ and $\{\eta_k\}$, and $i=0$.

3:  {\bf Loop}

4:  \quad Update the multipliers as 
$\lambda_k^{(i+1)} = \max\left\{\lambda_k^{(i)} +\eta_{\lambda_k} \dfrac{ \partial \mathcal{L}_{\text{P5} }}{\partial \lambda_k}, \quad 0\right\}, \;1\leq k \leq K.$

5:  \quad {\bf Initialize} $T_{c,k}^{(0)}$ and $t=0$.

6:  \quad {\bf Loop}

7:  \qquad $T_{c,k}^{(t+1)} = \max\left\{ T_{c,k}^{(t)} - \eta_k \dfrac{ \partial \mathcal{L}_{\text{P5} }}{\partial T_{c,k}^{(t)}},\; 0 \right\}$.

8:  \qquad $t=t+1$.

9:  \quad {\bf Until Convergence}

10: {\bf Until Convergence}

11: $\{T_{c,k}^* = T_{c,k},\;\forall k\}$ and calculate $T^*$.

12: Update the communication time $\{T_{c,k}\}$ using \eqref{Eq:CommunicationTimeUpdate}.

13: {\bf Output}: $\{T_{c,k}\}$.

\end{algorithm}

\subsubsection{Alternating Algorithm for Solving (P3)}
Based on Proposition \ref{Prop:MonotonicTimeAllocation}, the alternating optimization between Algorithms \ref{Alg:PowerQuantizationAllocation} and \ref{Alg:CommunicationTimeAllocation} leads to monotonically decreasing weighted distortion for (P3). Since (P3) is convex, the alternating method can optimally solve (P3), as summarized in Algorithm \ref{Alg:AlternatingMethod}, which suggests a linear convergence rate according to \cite{tran2019federated}.
\begin{algorithm}
\caption{Alternating Algorithm for Solving  (P3)}\label{Alg:AlternatingMethod}

1:  {\bf Input:}   Channel gains $\{H_{c,k}\}$ and auxiliary variables $y_{\ell,\ell^{'},n_k}$.
  
2:  {\bf Initialize} communication time $\{T_{c,k}\}$.

3:  {\bf Loop}

4:  \quad Solve sensing noise power $\{S_k\}$, quantization distortion $\{D_k\}$, and communication energy $\{E_{c,k}\}$ and discriminant gains $\left\{\mathcal{B}_{\ell,\ell^{'},n_k} \left(S_k,D_k\right)\right\}$, using Algorithm \ref{Alg:PowerQuantizationAllocation}.

5:  \quad Solve communication time $\{T_{c,k}\}$ using Algorithm \ref{Alg:CommunicationTimeAllocation}.

6:  {\bf Until Convergence}

7:  {\bf Output}: $\left\{\mathcal{B}_{\ell,\ell^{'},n_k} \left(S_k,D_k\right)\right\}$, $\{S_k\}$, $\{D_k\}$, $\{E_{c,k}\}$, and $\{T_{c,k}\}$.

\end{algorithm}

\subsection{Solution to (P2)}
Based on the previous results, (P3) can be optimally solved using the method of sum-or-ratios, together with the alternating algorithm in Algorithm \ref{Alg:AlternatingMethod}. The detailed procedure is summarized in Algorithm \ref{Alg:Sum-of-Ratios}. Then, by substituting the solution into the variable transformations in \eqref{Eq:VariablesTransform}, the optimal solution of (P2) can be obtained.
\begin{algorithm}
\caption{Sum-of-Ratios Based Optimal ISCC Scheme for Solving (P2)}\label{Alg:Sum-of-Ratios}

1:  {\bf Input:}   Channel gains $\{H_{c,k}\}$.
  
2:  {\bf Initialize} auxiliary variables $\{y_{\ell,\ell^{'},n_k}\}$.

3:  {\bf Loop}

4:  \quad Solve (P3) under given $\{y_{\ell,\ell^{'},n_k}\}$, using Algorithm \ref{Alg:AlternatingMethod}, and get   $\{S_k\}$, $\{D_k\}$, $\{E_{c,k}\}$, $\{T_{c,k}\}$, and $\left\{\mathcal{B}_{\ell,\ell^{'},n_k} \left(S_k,D_k\right)\right\}$.

5:  \quad Update the auxiliary variables as 
$y_{\ell,\ell^{'},n_k}  = \dfrac{1}{\mathcal{B}_{\ell,\ell^{'},n_k} \left(S_k,D_k\right)},\; \forall (\ell,\ell^{'},n_k),$

6:  {\bf Until Convergence}

7:  {\bf Output}: $\left\{\mathcal{B}_{\ell,\ell^{'},n_k} \left(S_k,D_k\right)\right\}$, $\{S_k\}$, $\{D_k\}$, $\{E_{c,k}\}$, and $\{T_{c,k}\}$.

\end{algorithm}

\section{Performance Evaluation}

\subsection{Experiment Setup}


\subsubsection{Communication model} In this experiment, we consider a network of $K=3$ ISAC devices, which are randomly located in a circular area of radius $50$ meters. The distance between the circle center and the AP is $450$ meters. The channel gain $H_{k}$ is modeled as $H_{k} = \left\vert \phi_{k}h_{k} \right\vert^{2}$, where $\phi_{k}$ and $h_{k}$ are the large-scale fading propagation coefficient and small-scale fading propagation coefficient, respectively. The large-scale propagation coefficient in dB from device $k$ to the edge server is modeled as $[\phi_{k}]_{\text{dB}}=-[\text{PL}_{k}]_{\text{dB}} +  [\zeta_{k}]_{\text{dB}}$, where $[\text{PL}_{k}]_{\text{dB}} = 128.1+37.6\log_{10}\text{dist}_{k}$ ($\text{dist}_{k}$ is the distance in kilometer) is the path loss in dB, and $[\zeta_{k}]_{\text{dB}}$ accounts for the shadowing in dB. In the simulation, $[\zeta_{k}]_{\text{dB}}$ is Gauss-distributed random variable with mean zero and variance $\sigma^{2}_{\zeta}$. The small-scale fading is assumed to be Rayleigh fading, i.e., $h_{k} \sim \mathcal{CN}(0,1)$. 

\subsubsection{Inference task} 
In our simulation, we apply the wireless sensing simulator in \cite{Li2021SPAWC} to simulate various high-fidelity human motions and generate human motion datasets. The inference task is to identify four different human motions, i.e., \textit{child walking}, \textit{child pacing}, \textit{adult walking}, and \textit{adult pacing} via the design of ISCC. Similar to the setup in \cite{matlab}, the heights of children and adults are assumed to be uniformly distributed in interval $[0.9\text{m}, 1.2\text{m}]$ and $[1.6\text{m}, 1.9\text{m}]$, respectively. The speed of standing, walking, and pacing are $0$ m/s, $0.5H$ m/s, and $0.25H$ m/s, respectively, where $H$ is the height value. The heading of the moving human is set to be uniformly distributed in $[–180^{\circ}, 180^{\circ}]$. 

\subsubsection{Inference model}
Two machine learning models, i.e., SVM and MLP neural network, are considered for inference in the experiments, respectively. The magnitudes of the feature elements are taken as the inputs of the learning models. The neural network model has 2 hidden layers with 80 and 40 neurons, respectively. Both models are trained on 800 data samples without any distortion, i.e., sensing clutter, sensing noise, and quantization distortion. The inference experiments for test accuracy are implemented over 200 data samples with distortion.

Unless specified otherwise, other simulation parameters are stated in Table \ref{tab:sensing-parameters}. All experiments are implemented using Python 3.8 on a Linux server with one NVIDIA\textsuperscript{\textregistered} GeForce\textsuperscript{\textregistered} RTX 3090 GPU 24GB and one Intel\textsuperscript{\textregistered} Xeon\textsuperscript{\textregistered} Gold 5218 CPU. 

\begin{table}[tt]
	\caption{Simulation Parameters }
	\label{tab:sensing-parameters}
	\centering
	\begin{tabular}{l l l l}
		\hline
		\bfseries Parameter & \bfseries Value & \bfseries Parameter &  \bfseries Value \\
		\hline\hline
		Number of ISAC devices, $K$ & $3$  & Sensing noise variance, $\sigma_{r}^{2}$ & 1 \\
		Clutter variance, $\sigma_{c,k}^{2}$ & 1, 0.1, 0.5 & Quantization variance, $\delta_{k}^{2}$ & 1\\
		Number of features after PCA, $N_{K}$ & 50 &	Number of classes, $L$ & 4 \\
		Permitted latency, $T$ & 1.85 s &		Energy threshold, $E_{k}$ & 0.15 Joule \\
		Computation time for each device, $T_{m,k}$ & 0.1s & Computation energy for each device, $E_{m,k}$ & 0.01 Joule \\ 
		Variance of shadow fading, $\sigma_{\zeta}^{2}$ & 8 dB & Communication channel noise power, $\delta_{c}^{2}$ & $10^{-12}$ W \\
		Bandwidth for communication, $B$ & 200 Hz &
		Bandwidth for sensing, $B_{s}$ & $10$ MHz \\
		Sensing carrier frequency, $f_{c}$ & $60$ GHz &
		Chirp duration, $T_{0}$ & $10$$\mu s$ \\
		Unit sensing time, $T_{r,k}$ & $0.5$ s &
		Sampling rate, $f_{s}$ & $10$ MHz \\
		\hline
	\end{tabular}
\end{table}

\subsection{Inference Algorithms}

For comparison, we consider four schemes as follows.
\begin{itemize}
	\item \textit{Power-aware allocation}: The sensing power is first allocated randomly and then the other parameters are allocated by the scheme in \textbf{Algorithm 4}.
	\item \textit{Time-aware allocation}: The communication time is firstly allocated equally and then the other parameters are allocated by the scheme in \textbf{Algorithm 4}.
	\item \textit{Quantization-aware allocation}: The quantization bits is first allocated as $16$ bits for each ISAC device and then the other parameters are allocated by the scheme in \textbf{Algorithm 4}.
	\item \textit{Optimal allocation (our proposal)}: All the parameters are allocated by the optimal ISCC scheme in \textbf{Algorithm 4}.
\end{itemize}

\subsection{Experimental Results}
In this part, the relations between the inference accuracy and discriminant gain regarding the two models are first presented. Then, the four algorithms are compared in terms of the SVM model and the neural network, respectively. Finally, the influence of number of participated devices on the inference accuracy is shown. 

\begin{figure}[t]
	\centering
	\includegraphics[width=0.5\textwidth]{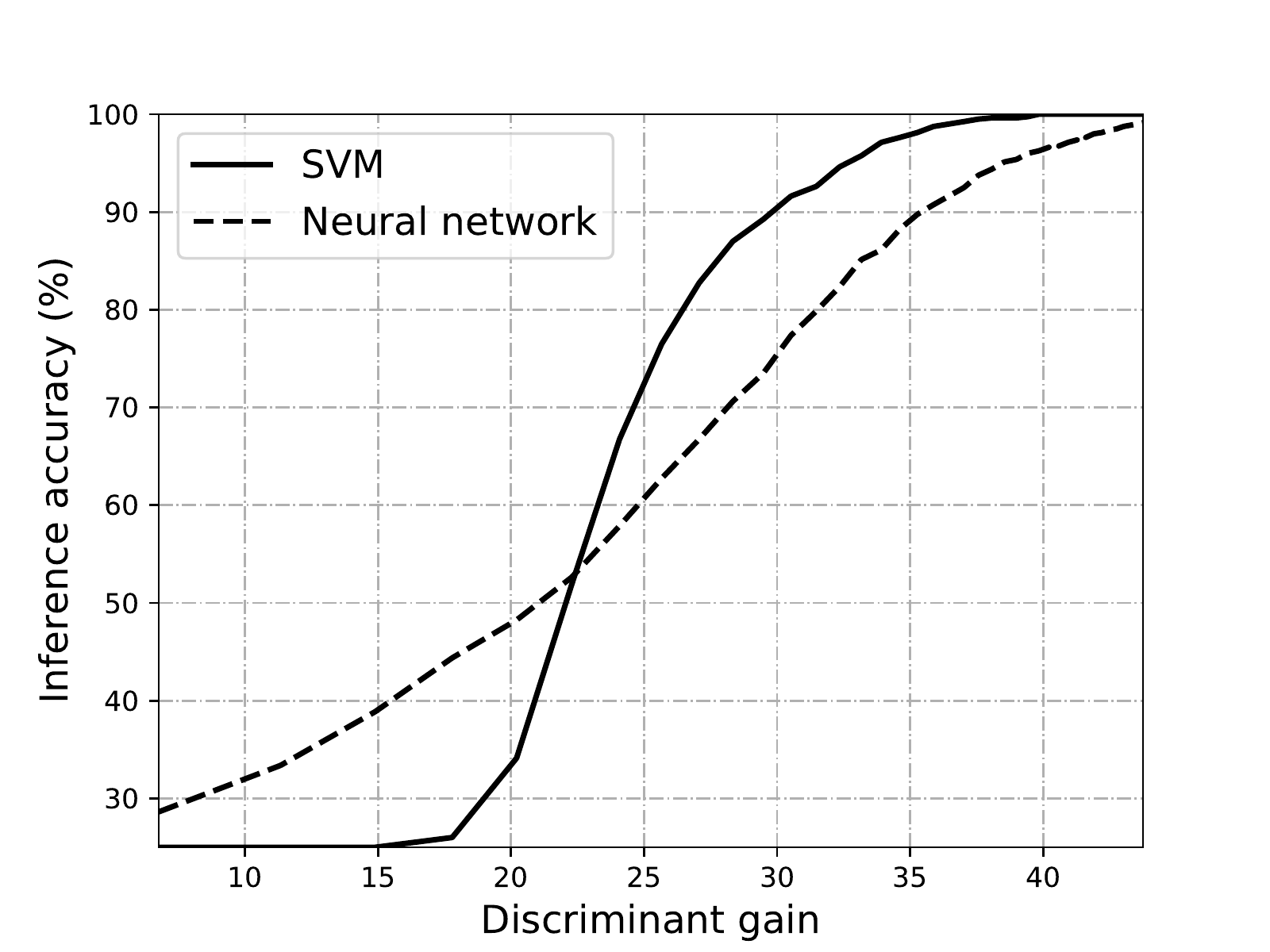}
	\caption{Inference accuracy versus discriminant gain.}
	\vspace{-0.5cm}
	\label{fig:disc_acc}
\end{figure}

\subsubsection{Inference accuracy v.s. discriminant gain}\label{Sect:Simulation1}
The relations between the inference accuracy and discriminant gain regarding the SVM model and the MLP neural network are shown in Fig. \ref{fig:disc_acc}. It is observed that the inference accuracy increases as the discriminant  gain grows for both models. Besides, when the discriminant gain is large, i.e., the distortion of the samples caused by sensing and quantization is small, the SVM outperforms the neural network. This is because the training of the neural network is overfitting as its model is complicated compared to its training dataset size. However, the neural network is more robust than the SVM when the discriminant gain is small, i.e., the distortion is large. It is also observed that when the discriminant gain is too large, the accuracy increases slowly because the centriods of different classes are too far apart in this case, and increasing the discriminant gain does not help much to increase the accuracy.

\begin{figure*}[t]
	\centering
	\subfloat[Inference accuracy with SVM versus energy threshold]{\includegraphics[width=0.495\textwidth]{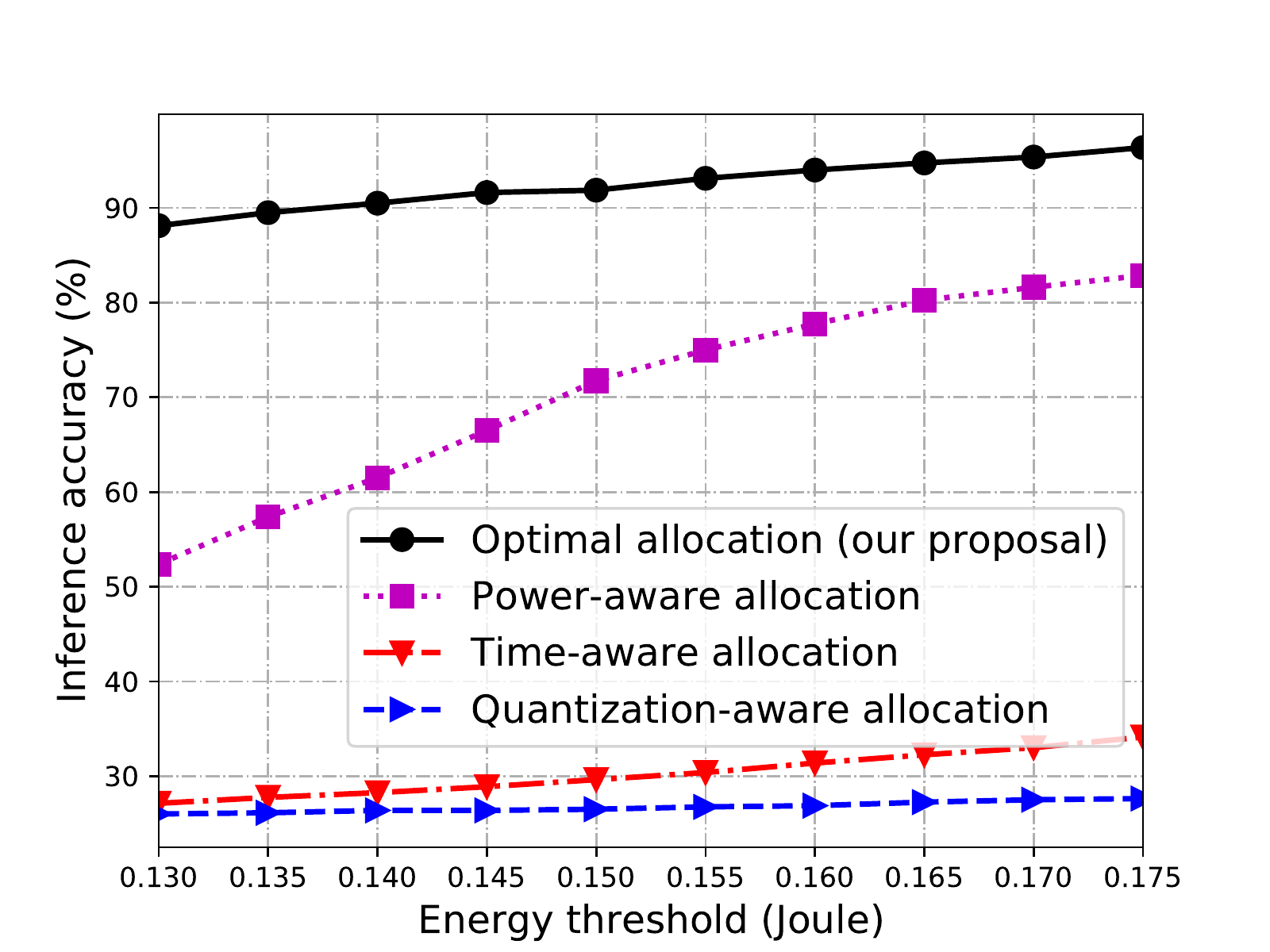}
		\label{fig:energy_accuracy_svm}}\hfil
	\subfloat[Inference accuracy with SVM versus permitted latency]{\includegraphics[width=0.485\textwidth]{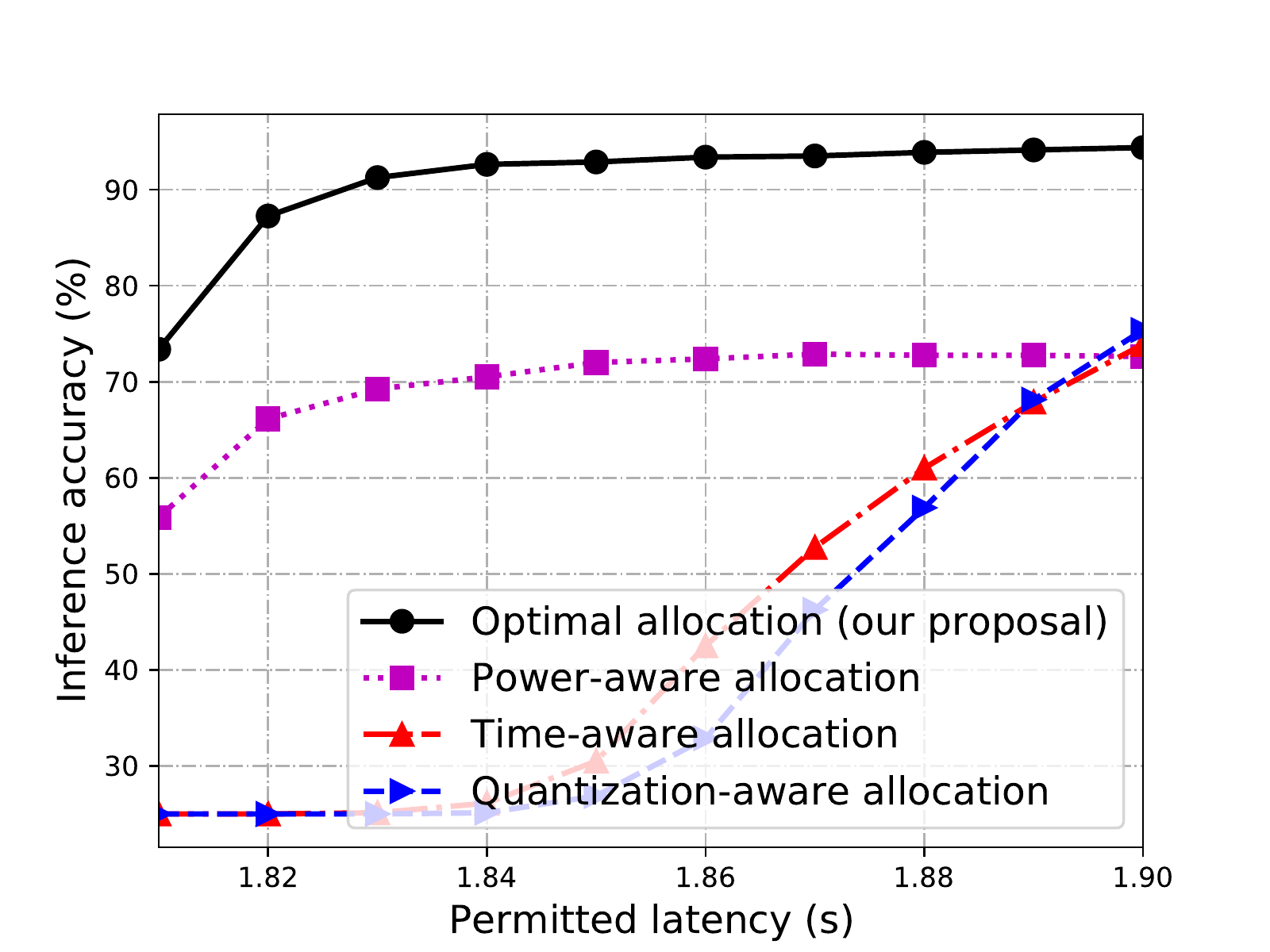}
		\label{fig:time_accuracy_svm}}
	\caption{Performance comparison of the SVM among different schemes.}
	\vspace{-0.5cm}
	\label{fig:accuracy_svm}
\end{figure*}

\subsubsection{Inference accuracy of SVM}
The inference accuracy of the SVM model is presented in Fig \ref{fig:accuracy_svm}. From the figure, the performance of all schemes increases as the resources, i.e., energy threshold of each device and the permitted latency for the inference task, increase. Besides, the proposed optimal allocation scheme outperforms the other three baseline schemes. Furthermore, in the case of long permitted latency, the performance of the power-aware allocation scheme remains unchanged as the permitted latency continuously increases. The reason is that the sensing noise is dominant in this case. 

\subsubsection{Inference accuracy of neural network}
\begin{figure*}[!t]
	\centering
	\subfloat[Inference accuracy  versus energy threshold]{\includegraphics[width=0.485\textwidth]{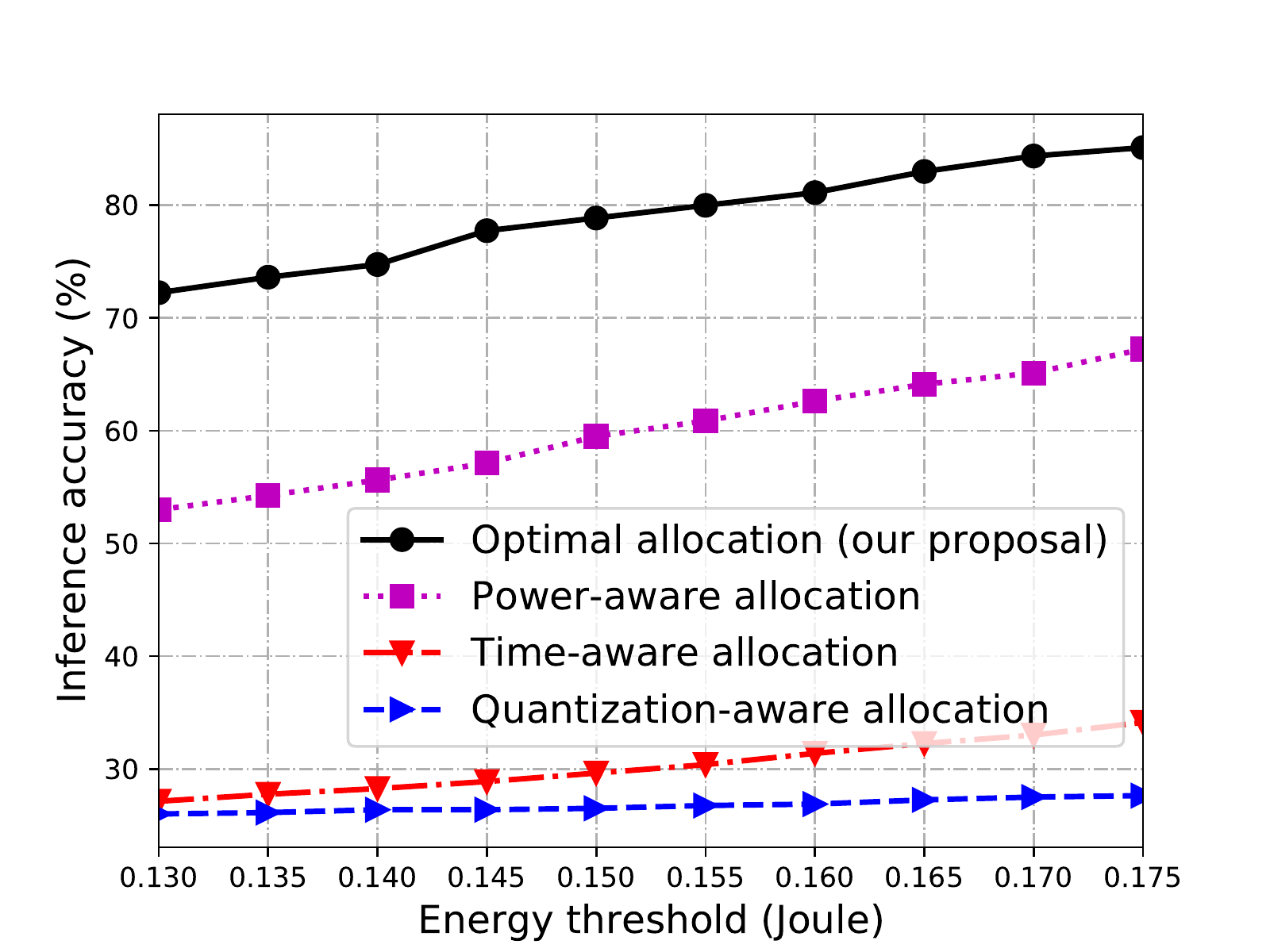}
		\label{fig:energy_accuracy_mlp}}\hfil
	\subfloat[Inference accuracy versus permitted latency]{\includegraphics[width=0.49\textwidth]{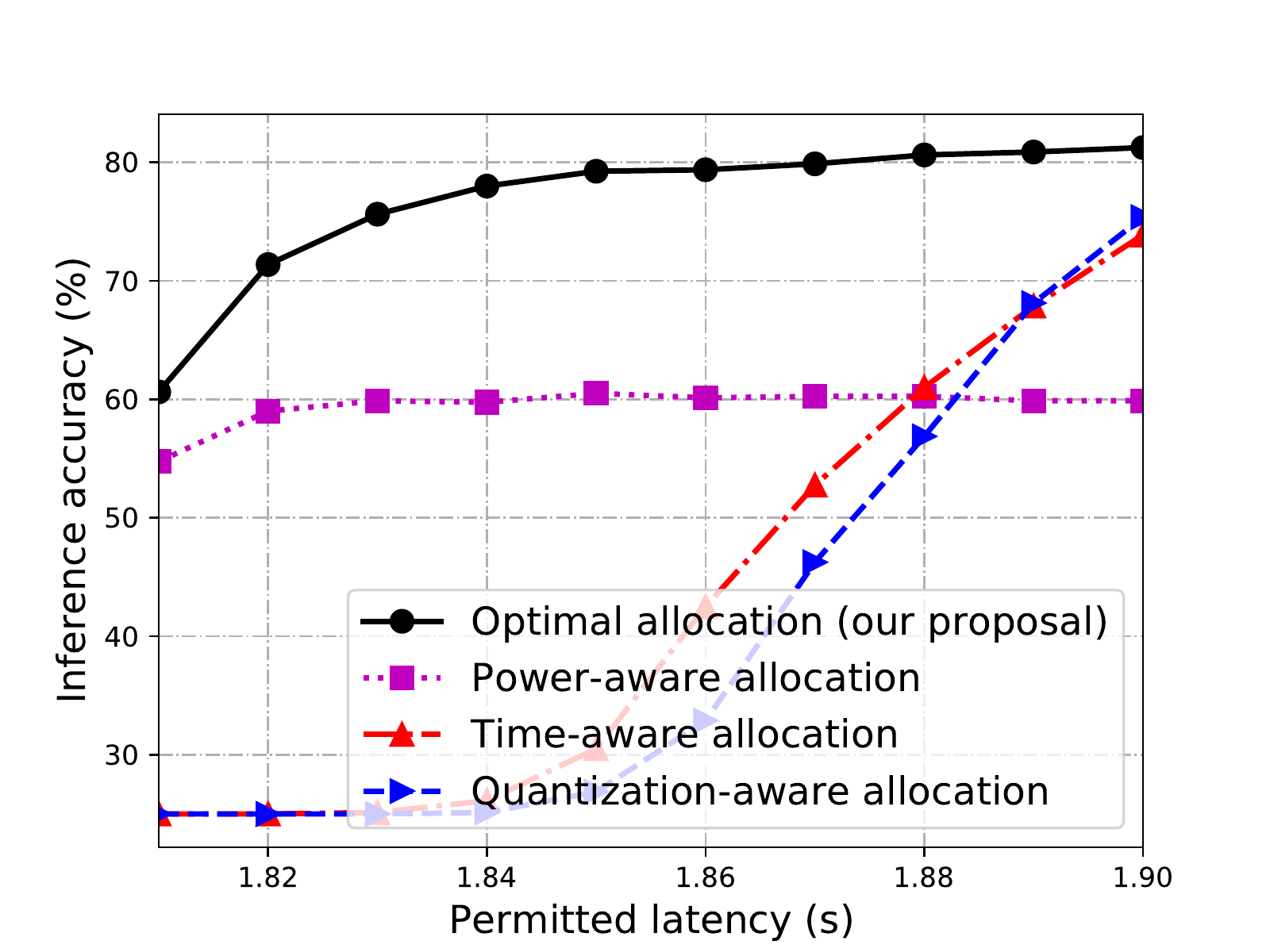}
		\label{fig:time_accuracy_mlp}}
	\caption{Performance comparison of the neural network  among different schemes.}
	\vspace{-0.5cm}
	\label{fig:accuracy_mlp}
\end{figure*}

The inference accuracy of the MLP neural network model in terms of the energy threshold and the permitted latency is shown in Fig. \ref{fig:accuracy_mlp}. Again, as more resources are allocated, the performance of all schemes increase. Besides, the proposed optimal allocation scheme achieves the best performance. Furthermore, the longer permitted latency will not lead to better performance for the power-aware allocation scheme when the latency is large, for a similar reason in the scenario of the SVM model.


\begin{figure}[t]
	\centering
	\includegraphics[width=0.5\textwidth]{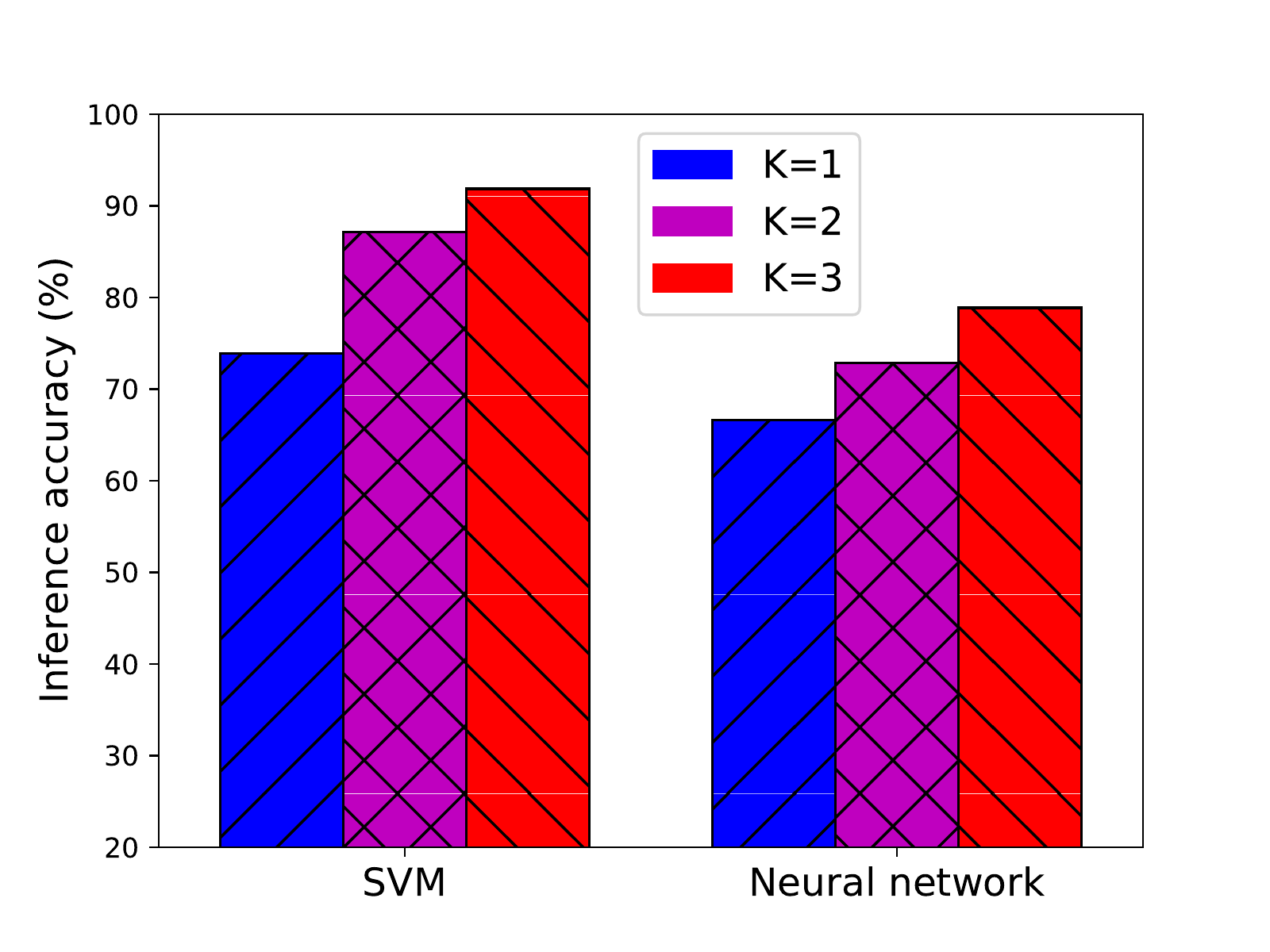}
	\caption{Inference accuracy comparison among different models under different number of ISAC devices.}
	\label{fig:num_radars}
\end{figure}

\subsubsection{Inference accuracy v.s. number of ISAC devices}
In Fig. \ref{fig:num_radars}, the inference accuracy of both models in terms of different number of ISAC devices are presented. For both cases, as the number of devices increases, better inference accuracy is achieved. The reason is that providing more features to the inference task can lead to a larger feature space, which can further make the distance, i.e., the discriminant gain, between arbitrary two different classes lager. In addition, the SVM outperforms the MLP, since the training of the neural network is overfitting.

The extensive experimental results above show that the proposed optimal ISCC scheme has the best performance and verify our theoretical analysis.

\section{Conclusion}

In this paper, we an optimal task-oriented ISCC scheme for edge AI inference. By jointly allocating the sensing and communication power, quantization bits, and communication time to maximize the discriminant gain of the received features, the accuracy is enhanced for real-time inference tasks.

This work opens several interesting directions for inference-task-oriented designs.  One is the ISAC device scheduling, i.e., the feature selection, for inference accuracy maximization when the radio resources, e.g., time and frequency bands, are scarce. Another is to enhance the inference accuracy in the broadband systems with frequency-selective wireless channels.

\appendix
\subsection{Proof of Lemma \ref{Lma:SumOfRatios}}\label{Apdx:LmaSumOfRatios}
The objective function of (P2) can be re-written as \eqref{Eq:DG11}, where
all $\{\mathcal{A}_{\ell,\ell^{'},n_k}\}$ are constants. In addition, $\{-\mathcal{B}_{\ell,\ell^{'},n_k}\left( S_k, D_k\right)\}$ for all $(\ell,\ell^{'},n_k)$ are linear. Obviously, each ratio is quasi-linear. Hence, (P2) can be optimally solved by the sum-of-ratios method if its feasible region is convex, according to \cite{jong2012efficient,yu2015joint}. In the next, we will show that the constraints are convex. The first constraint in (P2) is
 $   \sum\nolimits_{k=1}^{K} ( T_{r,k} + T_{m,k} + T_{c,k} ) \leq T,$ 
which forms a linear set and hence is convex. In the second constraint,  
\begin{equation}
     N_k \log_2\left( 1+\dfrac{1}{D_k} \right) \leq T_{c,k} B \log_2\left(1 + \dfrac{E_{c,k} H_{c,k}}{ T_{c,k} \delta_c^2 } \right), \; 1\leq k \leq K,
\end{equation}
the left part is convex, as its second derivative is positive.
The right part of the second constraint can be linearly transformed from $f(x,y) = x\log_2(1+y/x)$, which can be easily shown to be concave. 
As a linear transformation preserves convexity, the right part of the second constraint is a concave function. Thus, the second constraint forms a convex set. Next, the third constraint, i.e.,
 $ \sigma_r^2T_{r,k}/S_k +E_{m,k} + E_{c,k}\leq E_k$,  
also forms a convex set.  

\subsection{Proof of Proposition \ref{Prop:MonotonicTimeAllocation}}\label{Apdx:LmaMonotonicTimeAllocation}
In (P3), the second constraint is 
\begin{equation}
N_k\log_2\left( 1+\dfrac{1}{D_k} \right)\leq T_{c,k} B\log_2\left( 1+  \dfrac{ E_{c,k} H_{c,k} }{ T_{c,k} \delta_c^2 } \right), \; 1\leq k \leq K, 
\end{equation}
whose right-hand part is a strictly decreasing function of $T_{c,k}$. That is to say, with increasing $T_{c,k}$, smaller communication energy $E_{c,k}$ is used to satisfy this constraint for each device. Then, consider the final constraint in (P3), given as
$\left\{\sigma_r^2T_{r,k}/S_k + E_{m,k}+ E_{c,k}  \leq E_k,\;\; 1\leq k \leq K \right\},$ 
where smaller $E_{c,k}$ can lead to smaller sensing noise $S_k$. Next, according to $\mathcal{B}_{\ell,\ell^{'},n_k}\left( S_k, D_k\right)$ defined in \eqref{Eq:Fraction}, it is a linearly increasing function of $S_k$. Hence, the objective function of (P3) increases, which further leads to an enhanced discriminant gain according to \eqref{Eq:DG11}.

\bibliographystyle{IEEEtran}
\bibliography{reference}

\end{document}